\documentclass[aps,prc,preprint,superscriptaddress, nofootinbib]{revtex4-1}
\usepackage{amsmath}
\usepackage{graphicx}
\usepackage{units}
\usepackage{url}

\usepackage[dvipsnames]{xcolor}
\usepackage{color}
\definecolor{darkblue}{rgb}{0.0, 0.0, 0.55}
\definecolor{cite}{rgb}{0.0, 0.34, 0.25}
\definecolor{midgreen}{rgb}{0.52, 0.73, 0.4}

\begin{document}
\pdfminorversion=7
%\preprint{}

\title{Flow harmonic distributions and factorization breaking at RHIC}


%%%\author{}
%\thanks{}
%%%\affiliation{}
\author{Leonardo Barbosa}
\affiliation{Instituto de F\'{i}sica, Universidade de S\~ao Paulo, Rua do Mat\~ao 
1371,  05508-090 S\~ao Paulo-SP, Brazil}

\author{Fernando G. Gardim}
\affiliation{Instituto de Ci\^encia e Tecnologia, Universidade Federal de Alfenas, 37715-400 Po\c{c}os de Caldas-MG, Brazil}

\author{Fr\'ed\'erique Grassi}
\affiliation{Instituto de F\'{i}sica, Universidade de S\~ao Paulo, Rua do Mat\~ao 
  1371,  05508-090 S\~ao Paulo-SP, Brazil}

\author{Mauricio Hippert}
\affiliation{Illinois Center for Advanced Studies of the Universe, Department of Physics, University of Illinois at Urbana-Champaign, Urbana, IL 61801, USA}

\author{Pedro Ishida}
\affiliation{Instituto de F\'{i}sica, Universidade de S\~ao Paulo, Rua do Mat\~ao 
1371,  05508-090 S\~ao Paulo-SP, Brazil}

\author{Matthew Luzum}
\affiliation{Instituto de F\'{i}sica, Universidade de S\~ao Paulo, Rua do Mat\~ao 
1371,  05508-090 S\~ao Paulo-SP, Brazil}

\author{Meera V. Machado}
%\altaffiliation[Also at ]{Abzu ApS, Orient Plads 1, 1., 2150 Nordhavn, Denmark}
\affiliation{Abzu ApS, Orient Plads 1, 1., 2150 Nordhavn, Denmark}

\author{Jacquelyn Noronha-Hostler}
\affiliation{Illinois Center for Advanced Studies of the Universe, Department of Physics, University of Illinois at Urbana-Champaign, Urbana, IL 61801, USA}

\author{Christopher Plumberg}
\affiliation{Natural Science Division, Pepperdine University, Malibu, CA 90263, USA}
  

\date{\today}

\begin{abstract}
  Data obtained at RHIC can be reproduced with relativistic viscous hydrodynamic simulations by adjusting the viscosity and initial conditions but it is difficult to disentangle these quantities.
 % The problem should be even more complicated at lower energies relevant to explore the QCD phase diagram, in particular for the Beam Energy Scan.
  It is therefore important to find orthogonal observables to constrain the initial conditions separately from the viscosity.
  New observables have been measured at the LHC and  shown to be  sensitive to initial conditions and less to medium properties, specifically factorization breaking ratios appears to be promising.
  Here we consider two initial condition models, NeXus and TRENTO.
  %) that reproduce these quantities reasonably well at the LHC   and then we make predictions for RHIC. 
  While both models yield similar results for the scaled flow harmonic distributions both at the LHC and RHIC, they lead to  quantitatively much more  different patterns for the factorization breaking ratios at RHIC than  at LHC, due to the shorter lifetime. Additionally, 
  not only final state interactions but also initial free streaming matter in these ratio predictions and differences between these are enhanced at RHIC compared to LHC. Therefore  experimental factorization breaking ratios  at RHIC top energy (for transverse momentum)
  would be interesting to get.

\end{abstract}

%\keywords{}

\maketitle

\section{Introduction}

Relativistic heavy ion collisions are being performed at RHIC and  LHC to study the Quark Gluon Plasma. It was first discovered as a new state of matter created at high temperature and nearly zero net baryonic density. Now the challenge is to create it at non-zero net baryonic density, explore its phase diagram and  unravel its conjectured critical point \cite{Ratti:2018ksb,Bzdak:2019pkr,Dexheimer:2020zzs,Monnai:2021kgu}. Experimental facilities around the world are participating in this effort or planning to (BES-RHIC in the USA \cite{STARnote,Cebra:2014sxa}, HADES \cite{Galatyuk:2014vha} and FAIR \cite{Friese:2006dj,Tahir:2005zz,Lutz:2009ff,Durante:2019hzd} in Germany, NICA in Russia \cite{Kekelidze:2017tgp,Kekelidze:2016wkp}, J-PARC in Japan).

However, one of the largest uncertainties in the description of the Quark Gluon Plasma is the initial state because it cannot be directly probed by experiments.  Rather, experiments only measure the after math of a heavy-ion collisions, following the pre-equilibrium hydrodynamic state, hydrodynamics, and hadronic interactions \cite{reviewHeinz,reviewGale,reviewdeSouza,reviewQGP5}.  Thus, one must disentangle different contributions to the final measured flow observables that come from the medium itself vs. the initial state.  This is even more challenging at low beam energies where finite baryonic potential effects begin to play a role.

Many data can be reproduced by adjusting suitably the fluid viscosities and its initial conditions and it is difficult to disentangle these quantities \cite{reviewHeinz}. However, in recent years it has become clear that some quantities are fairly independent of the fluid transport  properties and reflect fluctuations in its initial state \cite{Giacalone:2017uqx}. This is the case (as detailed below) of the scaled harmonic flow distributions and flow factorization ratios  which have been measured at the LHC. In fact these data provide a rather strong test of initial condition.
Therefore, it would be important to have such measurements made at RHIC at least at the highest energy where the equation of state is known from lattice simulations and there are fewer competing effects.
 Since these measurements have not been made yet, one can test if a model has correct fluctuations by comparing with harmonic flow distributions \cite{ATLAS12,ATLAS13,ALICE13}
and flow factorization  data \cite{Khachatryan:2015oea, Acharya:2017ino} obtained at LHC energy and then make predictions for RHIC top energy. In this paper, we illustrate this with two initial condition models:
 NeXus  \cite{NeXus} and TRENTO \cite{Moreland:2014oya}.

 Here we systematically study the beam energy scaling of the flow distributions and factorization breaking and identify specific centralities and kinematic cuts that are the most promising for distinguishing initial state models. Then we 
 discuss the differences in predictions and their origin for top RHIC energies. We also examine the effect of free streaming and final state interactions.
 
\section{Relevant experimental observables}\label{sec:obs}   

\subsection{Flow harmonic distributions}

In this work our primary focus will be on the flow harmonics that can be measured at RHIC. In order to calculate them from our theoretical models we  expand the particle azimuthal distribution as a Fourier series:

\begin{equation}
\frac{dN}{d\phi}=\frac{N}{2\pi}\left[1+2\sum\limits_{n=1}^{\infty}v_n \cos[n(\phi-\Psi_n)]\right],
\end{equation}
with
\begin{equation}
V_n \equiv v_n e^{in\Psi_n}=\frac{\int d\phi e^{in\phi}\frac{dN}{d\phi}}
{\int d\phi \frac{dN}{d\phi}}
\end{equation}
where $V_n$ is the flow vector and $\phi$ is the particle transverse momentum azimuthal angle. The magnitude and orientation of the flow vectors, $v_n$ and $\Psi_n$, vary from an event to another due to quantum fluctuations in the initial conditions.

Because of the event-by-event fluctuations that occur in the experiment, hydrodynamic models must also incorporate these same fluctuations in order to reproduce experimental results. Thus, one runs a large set of initial conditions (events) through relativistic hydrodynamic simulations, each which produces its own flow harmonic $v_n$ for that specific event. Then, one can construct a $v_n$ distribution $P(v_n)$ from this ensemble of events, which can be directly compared to unfolded distributions from  ATLAS \cite{ATLAS12,ATLAS13} and ALICE \cite{ALICE13} experimental data.

In fact, these distributions offer a stringent test of initial conditions. 
Many models fail to reproduce these observables.
This is the case \cite{ATLAS12,ATLAS13} of  Glauber  \cite{glauber1,glauber2,glauber3} and MC-KLN \cite{MCKLN} initial conditions   (variations have also been studied \cite{Renk14,Ghosh16}). The following models  provide satisfactory results:  IP-Glasma 
 \cite{Schenke12,McDonald:2016vlt} as shown in  \cite{Schenke13}, EKRT     \cite{EKRT} as studied in \cite{Niemi15}, AMPT \cite{Zhang:1999bd} and TRENTO \cite{Moreland:2014oya} as discussed in \cite{Zhao:2017yhj}.  Below we show that NeXus also leads to reasonable results.

 The scaled harmonic distributions $P(v_n/\langle v_n \rangle)$ provide information about the amount of fluctuations, and are known to be relatively independent of viscosity \cite{Niemi12,Niemi15} (and small scale structure \cite{Gardim:2017ruc,Renk14}),
 which can be understood as follows. For a given centrality class, where the average value of  $v_n$ is $\langle v_n \rangle$,  
 the initial conditions drive the magnitude of  $v_n$ compared to $\langle v_n\rangle$ (so do not cancel in $v_n/\langle v_n \rangle$)
 while the fluid properties are approximately similar for a given event and the average of events (so should cancel in  $v_n/\langle v_n \rangle$).
Thus, the width of the $P(v_n/\langle v_n \rangle)$ distributions depends on having the right amount of fluctuations in the initial conditions.

A more precise understanding of the relationship between  $P(v_n/\langle v_n\rangle)$ and the initial conditions
can be obtained by defining the event eccentricities:
\begin{equation}
\epsilon_{m,n} e^{in\Phi_{m,n}}=-\frac{\int r dr d\phi r^m e^{i n \phi} \rho(r,\phi)}
{ \int r dr d\phi r^m  \rho(r,\phi)}
\end{equation}
where $r$ and $\phi$ are the spatial radius and azimuthal angle. When $m=n$, the simplified notations $\epsilon_n$ and $\Phi_n$ are used, which we will consider here. For $n=2$ or 3, it has been shown that in a given centrality window, $v_n \propto \epsilon_n$ holds not just in average \cite{sph1,sph2} but approximately for each event \cite{Niemi12,Niemi15,Fu15}. One expects therefore $P(v_n/\langle v_n\rangle)\sim P(\epsilon_n/ \langle\epsilon_n\rangle)$ for $n=2$ and 3 and this can even been shown for high $p_T$ and heavy flavor \cite{Noronha-Hostler:2016eow,Betz:2016ayq,Prado:2016szr,Katz:2019fkc}. Deviations from this are expected e.g. $v_2$  departs from  linear growth with $\epsilon_2$ for non-central collisions \cite{Niemi12,Niemi15,Fu15,Jaki15}, small systems \cite{Sievert:2019zjr}, and low beam energies \cite{Rao:2019vgy}. For $n>3$, the simple relation $v_n \propto \epsilon_n$ also breaks. For instance, $v_4$ is expected to be linear in $\epsilon_4$ for central collisions while it has an  $\epsilon_2^2$ behavior for non-central ones (see e.g.  \cite{sph1,sph2}),
so the relation $P(v_4/\langle v_4\rangle)\sim P(\epsilon_4/ \langle\epsilon_4\rangle)$ might hold but only for central collisions. Scalings such as  $v_n\propto \epsilon_{m,n}$ were studied in \cite{Niemi15,Fu15,Hippert:2020kde} and analytical approximations for the eccentricity distributions were proposed in  \cite{Voloshin:2007pc,JYPRL,JYPRC,JYPLB}.
The effect of temperature dependent viscosities was investigated in \cite{Gardim:2020mmy}.

Modern hydro codes include not only viscous hydrodynamics but also final state interactions (via a hadronic cascade model)
and some pre-thermalization stage (often free-streaming) . Because of the scaling $v_n \propto \epsilon_n$, we also expect that  $P(v_n/\langle v_n\rangle)$ are relatively independent of these two effects but we  check this below.

\subsection{Flow factorization  ratios}\label{sec:rfac}
 
Above we considered momentum integrated observables, however more detailed information can be obtained from differential quantities, such as the Pearson correlation between flow vectors in different momentum range known as the flow factorization ratio  \cite{gardim13}:

\begin{equation}
  r_n(p_1,p_2)=\frac{ V_{n\Delta}(p_1,p_2)}{\sqrt{ V_{n\Delta}(p_1,p_1) V_{n\Delta}(p_2,p_2)}}
  \label{eq:ratio}
 \end{equation} 
where the   $V_{n\Delta}(p_1,p_2)$ are  the Fourier coefficients of the averaged  pair correlation $dN_{pairs}/d^3p_1 d^3p_2$. It can be re-written as:
\begin{equation}
 r_n(p_1,p_2)=\frac{\langle v_n(p_1)v_n(p_2)\cos\,n(\psi_n(p_1)-\psi_n(p_2))\rangle}
{\sqrt{\langle v_n^2(p_1) \rangle \langle v_n^2(p_2) \rangle}}
 .
\end{equation}
In hydrodynamics $r_n$ is  $\leq  1$  because of the fluctuations in the event-plane angles $\psi_n(p_T)$ and in the harmonic flow magnitudes  $v_n(p_T)$ ($\langle v_n(p_1)v_n(p_2) \rangle \neq 
\sqrt{\langle v_n^2(p_1) \rangle \langle v_n^2(p_2) \rangle}$), and as also happened with scaled flow, factorization breaking is not very sensitive 
\footnote{In \cite{Khachatryan:2015oea}, fig.3 with  values of shear viscosity from 0.08 to as large as 0.20,
the $r_2$ results are affected by less than 7\% for MC Glauber initial conditions and less than  3 \% for  MC KLN initial conditions.
In \cite{kozlov14}, fig. 13 and 14 with  values of shear viscosity between 0. and 0.08, 
for  MC Glauber initial conditions the $r_2$ and $r_3$ results are affected by less than 1\%.
In addition, the effect of viscosity decreases for less central collisions.} to viscosity
\cite{kozlov14,heinz15,Khachatryan:2015oea}.

It is known that final state interactions affect the factorization breaking ratio \cite{McDonald:2016vlt}, we will see that free streaming can have an even stronger effect.

Data for these ratios  were obtained at the LHC by CMS \cite{Khachatryan:2015oea} and ALICE \cite{Acharya:2017ino}, providing another strong test of initial condition models. It was observed in  \cite{Khachatryan:2015oea,heinz15} that neither MC-Glauber nor MC-KLN was compatible with data in the whole range of $p_T$ intervals and centralities. Though leading to reasonable results for $r_2$, AMPT, TRENTO and IP-Glasma predict a drop in $r_3$ for large $p_T$  differences stronger than  in data \cite{Zhao:2017yhj,McDonald:2016vlt}.

\section{Heavy-Ion simulations}

\subsection{Initial conditions}\label{sec:IC}

Here we compare two different initial conditions models relevant for both LHC (Pb-Pb 2.76 TeV) and RHIC (Au-Au 200 GeV) energies. NeXus \cite{Werner:1993uh,NeXus,Drescher:2000ec} is based on a partonic string model inspired on Gribov-Regge theory. 
Among models that have initial nucleon-size fluctuations, NeXus has the advantage to have been thoroughly tested at top RHIC energy and provides a coherent picture of data \cite{NeXspectra,Andrade06,Andrade08a,Andrade08b,Gardim11,Gardim12a,Takahashi09,Qian12}, and has also been compared to some of the LHC data \cite{diss_meera,Gardim:2020fxx}. 

We will then contrast NeXus with the TRENTO initial condition framework that is a phenomenological model that parameterizes the initial conditions.  For TRENTO initial conditions, we will use two different parameter sets that are both motivated by different Bayesian analyses.  The first parameterization of TRENTO was based on the Bayesian analysis from \cite{Moreland:2014oya,Bernhard:2016tnd} when no free-streaming was considered. For this case, we use the central values from their Bayesian analysis such that  $p=0$, $k=1.6$ and $w=0.51$ fm. This parameters set has been shown to fit to charged particle yields, $\langle p_T \rangle$, and event-by-event flow fluctuations \cite{Alba:2017hhe,Giacalone:2017uqx,Nijs:2020roc}. Additionally, many other theoretical predictions exist from TRENTO that remain to be confirmed \cite{Summerfield:2021oex,Ke:2016jrd,Mordasini:2019hut,Giacalone:2020lbm}. However, we note that correlations of $\langle p_T\rangle$ and $v_n$ fail to reproduce experimental data \cite{Giacalone:2020dln,ATLAS:2021kty}, it is not clear if this deviations arises from the initial state or hydrodynamic response.  

The second parameterization of TRENTO was also motivated by a Bayesian analysis but this time free-streaming was used \cite{Bernhard:2019bmu}. 
We note in particular that in this case,  the free streaming time is  $\tau_{fs}=1.16$ fm, $p\sim 0$,  $k=1.2$ but the nucleon width is $w=0.956$.
For the scaled harmonic distributions, we expect that $w$ has little effect (because it has an approximately similarly effect on $\epsilon_n$ and $\langle \epsilon_n\rangle$ \cite{kozlov14,Gardim:2017ruc}). We have checked (not shown) that this is indeed the case.
For the factorization breaking ratio, it was shown in 
\cite{kozlov14,Gardim:2017ruc} that the size of the hot spots (in particular here $w$) matters.

\subsection{Relativistic hydrodynamics}

For our three different sets of initial conditions, we apply different hydrodynamic approaches that have their respective parameters (e.g. transport coefficients and freeze-out temperatures) tuned to reproduce experimental spectra and flow harmonics.  The initial condition NeXus is used in the ideal 3+1 hydrodynamic code NeXSPheRIO (NeXus+SPheRIO) that was shown to reproduce experimental data in 
\cite{NeXspectra,Andrade06,Andrade08a,Andrade08b,Gardim11,Gardim12a,Takahashi09,Qian12}.  The initial condition of TRENTO without free-streaming was used in the 2+1 v-USPhydro that was shown to reproduce experimental data in \cite{Alba:2017hhe,Sievert:2019zjr,Rao:2019vgy}. The initial condition of TRENTO with free-streaming was used in the 
(2+1) VISHNU code that was shown to reproduce experimental data for example in \cite{Bernhard:2019bmu}. After the 2+1 or 3+1 relativistic hydrodynamic simulations, each is followed by a hadronic phase (without or with transport).  Following the hadronic phase then one can reconstruct the particle spectra (in this paper we consider only all charged particles). After the particle spectra is obtained, one can calculate the relevant experimental observables following the description in Sec.\ \ref{sec:obs}.

Each initial condition model requires specific parameters in the hydrodynamic simulation in order to reproduce experimental data.  Thus, for simplicity's sake, we couple each individual initial condition to the respective hydrodynamic model that they are commonly coupled to. 

In this paper we  consider NeXSPHeRIO (NeXus+SPheRIO) and Trento+v-USPhydro. We will describe each respective set-up below. However, we remark that both codes use the same numerical solver known as  Smoothed Particle Hydrodynamics, which is a Lagrangian method \cite{testNeX} to solve the equations of motion. There are two crucial differences between SPHeRIO and v-USPhydro, which are the dimensionality and out-of-equilibrium effects:  SPHeRIO is a 3+1 ideal hydrodynamic code whereas v-USPhydro is a 2+1 viscous hydrodynamic code. 

First we discuss the 3+1 ideal hydrodynamic framework of  NeXSPheRIO \cite{testNeX}. 
The equation of state has a critical point introduced phenomenologically \cite{eos}.
A Cooper-Frye freeze out is used with temperatures adjusted
to reproduce  observables for each centrality window.
Tests of this code
 against known solutions can be found in \cite{testNeX}.

 In contrast,   v-USPhydro \cite{vuspb,vuspbs} can incorporate both shear and bulk viscosity
 in 2+1 dimensions where longitudinal boost invariance is assumed. For simplicity, only 
  shear viscosity is considered where $\eta/s\sim 0.05$ and assumed constant.  
 TRENTO initial conditions are used.
The most state-of-the-art Lattice QCD equation of state \cite{Alba:2017hhe} at $\mu_B=0$ coupled to the particle list PDG16+ \cite{Alba:2017mqu} that includes all *-**** states from the Particle Data group. 
A constant temperature $T_{FO}=150$ MeV Cooper-Frye freeze out is employed.
Tests of this code
against known solutions were presented in \cite{vuspb,vuspbs}.

NeXSPheRIO and v-USPhydro both have no free streaming and a simple freeze out for particle emission, which  allows a better handle on the effects of the initial condition fluctuations.
For completeness, calculations were made using TRENTO initial conditions,
with the code VISHNU  (precisely  the implementation   \cite{Bernhard:2019bmu}).
 It is based on VISH2+1, a more traditional grid-based hydrodynamics code \cite{Song_2008} and has  a switching to  the hadronic cascade UrQMD at
 $T_{switch}=151$ MeV,
 instead of freeze out  \cite{Song_2011}.
It incorporates both shear and bulk viscosity in 2+1 dimensions (with boost invariance) and has
free streaming before the hydrodynamic phase.
The equation of state is a matching of lattice QCD results with a hadron gas calculation \cite{Bernhard:2019bmu}. This simulation leads to a good description of various data at LHC (yields of charged particles, transverse energy and identified particles, mean transverse momenta of these particles, anisotropic flow cumulants  of different orders, mean transverse momentum fluctuations).

In Table \ref{tab:modelsum}, a summary of the models used and corresponding parameters is presented.

\begin{table}
  \begin{tabular}{|ll|}
  \hline
  NeXSPheRIO & Perfect fluid  \\
  3+1         &     $T_{fo}$ adjusted for each centrality\\
   \hline
   TRENTO  & $p=0, k=1.6, w=0.51$\\
    2+1 & $\eta/s=0.05, T_{fo}=150$ MeV\\
    \hline 
  fTRENTO  &  $p=0, k=1.2, w=0.956$\\
  2+1  &  $  \tau_{fs}=1.16\,fm,  \eta/s(T), \zeta/s(T)$  \\
   \hline
  fTRENTO+UrQMD & as above with hadronic transport\\
    &   $T_{sw}=151$ MeV\\
     \hline
 \end{tabular}
 \caption{Summary of the models used and corresponding sets of parameters.
 NeXSPheRIO means NeXus+SPheRIO. TRENTO is TRENTO+v-USPhedro.
  fTRENTO stands for TRENTO initial conditions with free streaming and VISHNU 
 without UrQMD. fTRENTO+UrQMD is the same but with UrQMD.
 }
\label{tab:modelsum}
\end{table}

\section{Results}

\begin{figure*}[!ht]
  \includegraphics[width=0.8\textwidth]{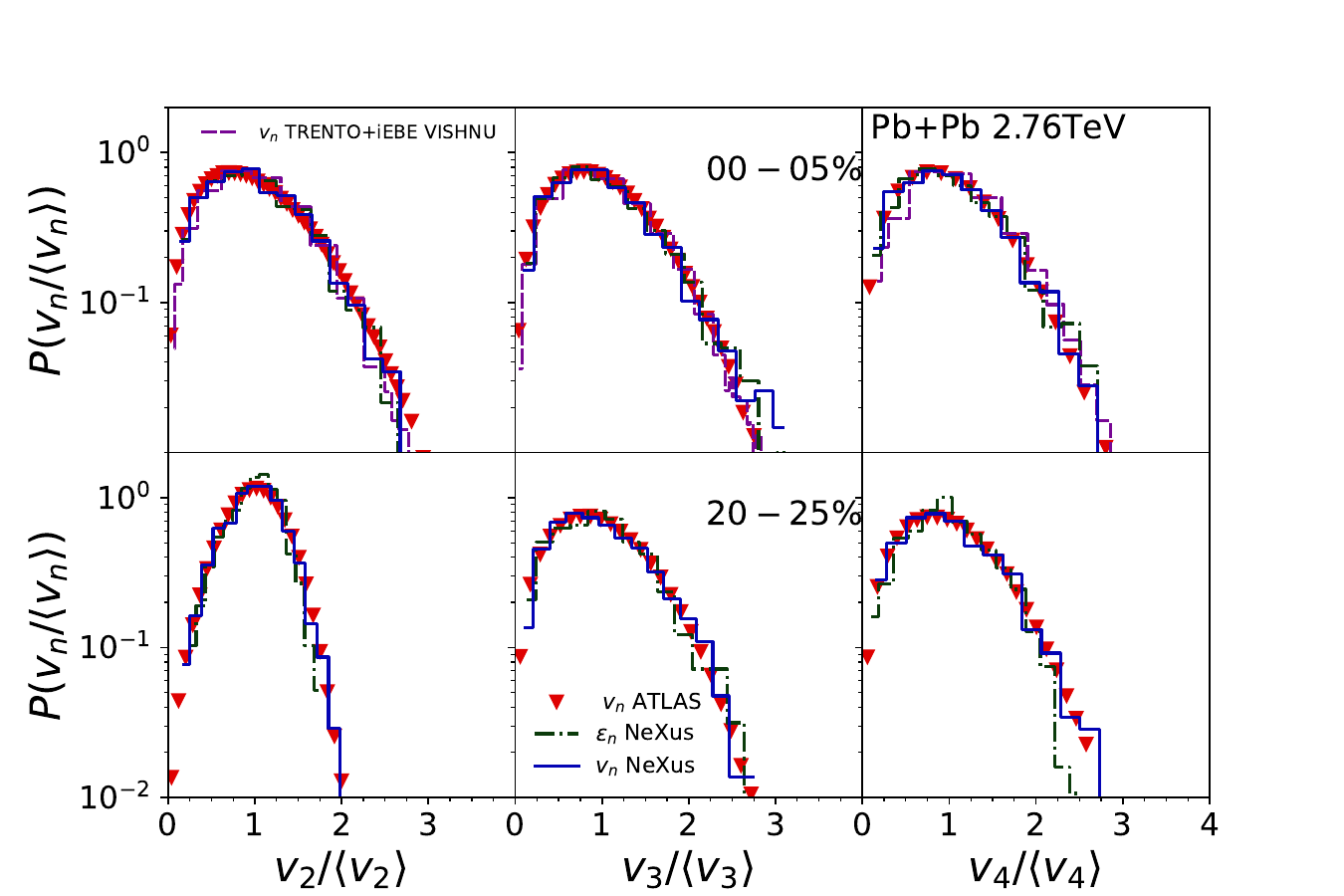}\\
   \includegraphics[width=0.8\textwidth]{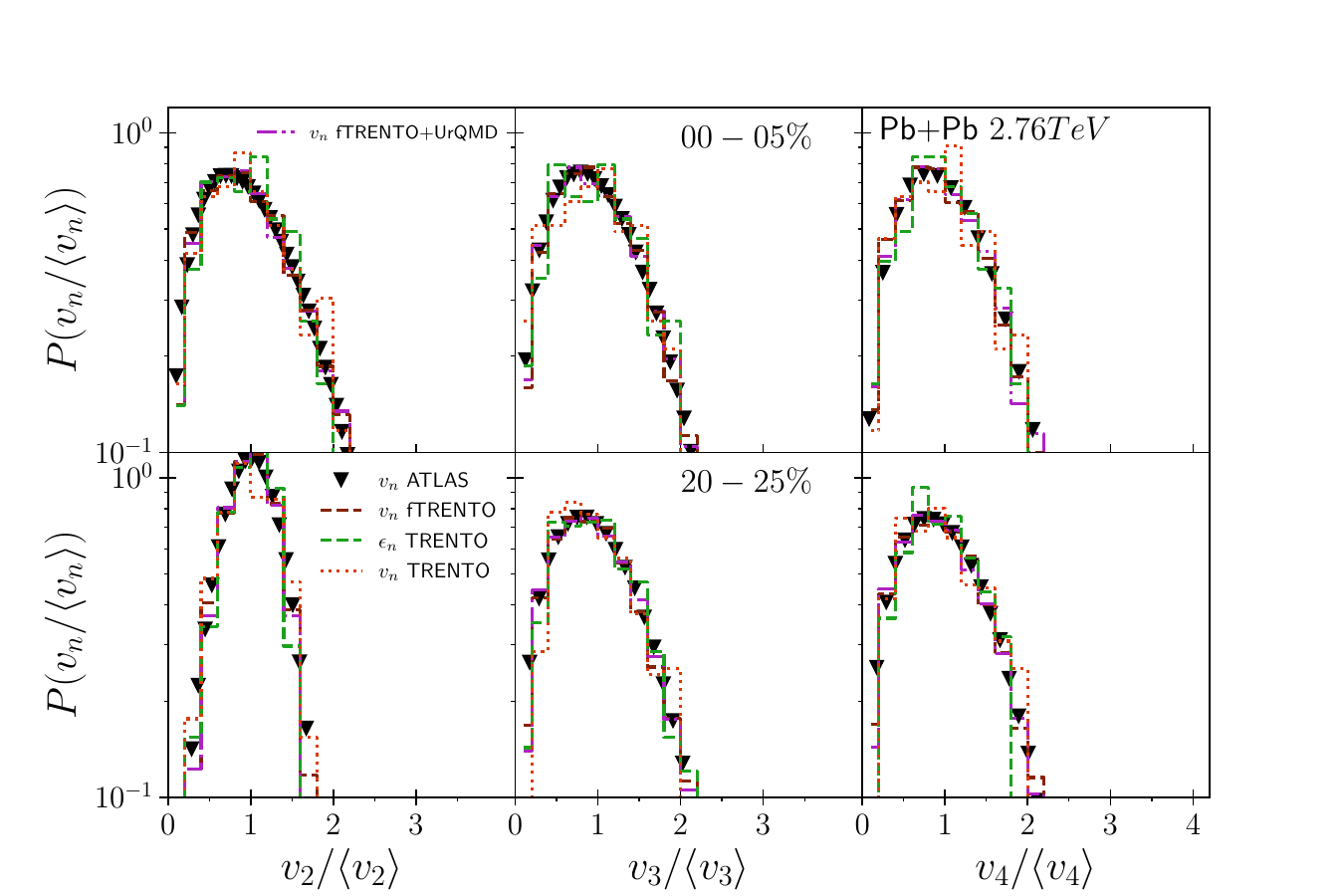}
\caption{Top: Comparison of scaled $\epsilon_n$ and $v_n$ distributions from NeXSPheRIO ($|\eta|<2.5$) with ATLAS data ($|\eta|<2.5$) \cite{ATLAS13} in the range
   $p_t>$ 0.5 GeV for Pb+Pb collisions at 2.76 TeV. (As expected the  right side tail  of the scaled $\epsilon_n$ may be  narrower than the  $v_n$'s one.)  Results from \cite{Zhao:2017yhj} have been added here for clarity.\\
  Bottom: Same comparison for TRENTO, fTRENTO and 
  fTRENTO+UrQMD.
}
\label{fig:NeXusscaleddist}
\end{figure*}

First we check how these models perform for the ATLAS scaled flow harmonic distributions at PbPb 2.76 TeV \cite{ATLAS13} for $v_2$, $v_3$, and $v_4$.
Fig. \ref{fig:NeXusscaleddist} top, shows that results coming from NeXus  initial conditions  agree with data.
  Therefore,
NeXus has the right amount of fluctuations in the initial conditions, since scaled flow distributions approximately
follow scaled
 eccentricity distributions. As expected, the eccentricity and flow harmonic distributions are nearly the same for these centralities (0-5\% and 20-25\%).  The one exception is for $v_4$ in the 20-25\% bin, which is unsurprising since $v_4$ experiences both a linear and non-linear response (from mixing with $v_2$).  In fact, reproducing the sign change of $v_4\left\{4\right\}^4$ is an open problem in the field \cite{Giacalone:2016mdr}. 

 In Fig. \ref{fig:NeXusscaleddist} bottom, results are shown for the TRENTO initial conditions with v-USPhydro (TRENTO),  as well as TRENTO with
 free streaming and  VISHNU but no UrQMD (fTRENTO)
  and TRENTO with
 free streaming and  VISHNU with UrQMD 
  (fTRENTO+UrQMD).
To not overcharge this figure, results from a different group \cite{Zhao:2017yhj} for TRENTO with VISHNU  and no free streaming (same parametrization as our case TRENTO)
  are shown in the top panel. 
All results agree with data. This can be understood with the same argument that was put forward to explain the independence on viscosity: free streaming and hadronic cascade affect approximately in the same way a given event and the average of events (in a centrality bin), therefore approximately cancel in $v_n/\langle v_n \rangle$.

\begin{figure*}[!ht]

  \includegraphics[width=0.8\textwidth]{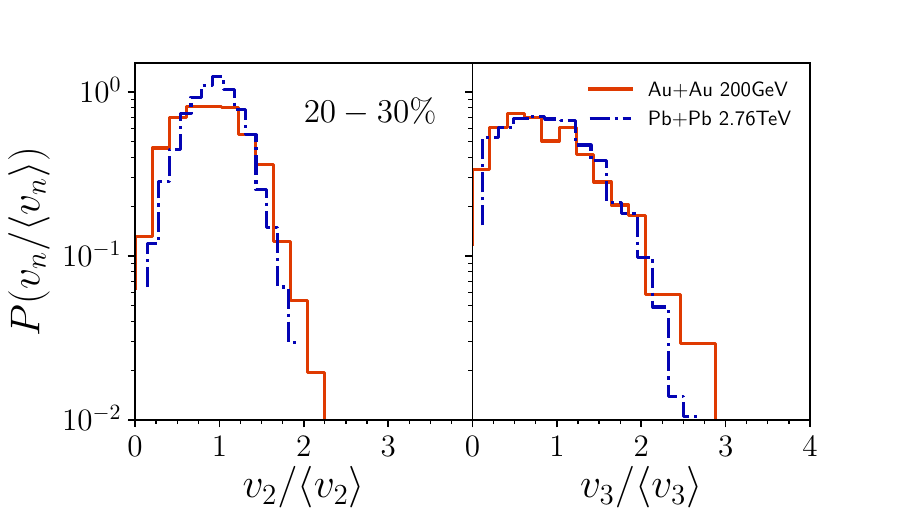}\\
     \includegraphics[width=0.8\textwidth]{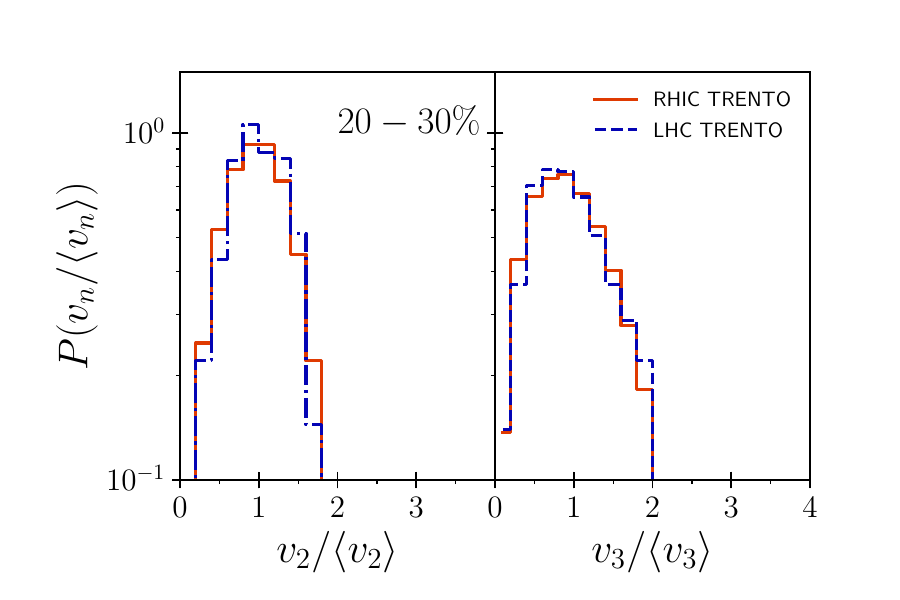}
\caption{Comparison of scaled  $v_n$ distributions for NeXus (top) and TRENTO (bottom)  at LHC  and RHIC  energies
  for the 20-30\% centrality window.}
\label{fig:NeXusdistenergydep}
\end{figure*}

We now turn to the energy dependence of these distributions: the scaled eccentricity distributions are not expected to depend strongly on the beam energy, so the same should hold for the scaled $v_n$ distributions.
Predictions for $P(v_n/\langle v_n \rangle)$  at 5.02 A TeV as well as  data and comparison at PbPb 2.76 TeV were presented  for IP-Glasma+MUSIC+UrQMD \cite{McDonald:2016vlt}, AMPT+VISHNU and TRENTO+VISHNU \cite{Zhao:2017yhj}. No difference was observed between these two LHC energies in all cases. In figure \ref{fig:NeXusdistenergydep}, NeXSPheRIO $v_n$ distributions are shown for LHC and top RHIC energies. They are also fairly similar even though the energies are more different.  From Fig.\ \ref{fig:NeXusdistenergydep} we see that RHIC has slightly larger fluctuations than LHC for both $v_2$ and $v_3$, this is consistent with \cite{Alba:2017hhe, Rao:2019vgy} which showed that $v_2\left\{4\right\}/v_2\left\{2\right\}$ is smaller at RHIC than the LHC (implying that fluctuations are larger at RHIC). In the same figure,
results are also shown for  TRENTO (fTRENTO and fTRENTO+UrQMD do not bring noticeable modifications), the difference between RHIC and LHC is very small.

\begin{figure*}[!ht]
  \includegraphics[width=0.8\textwidth]{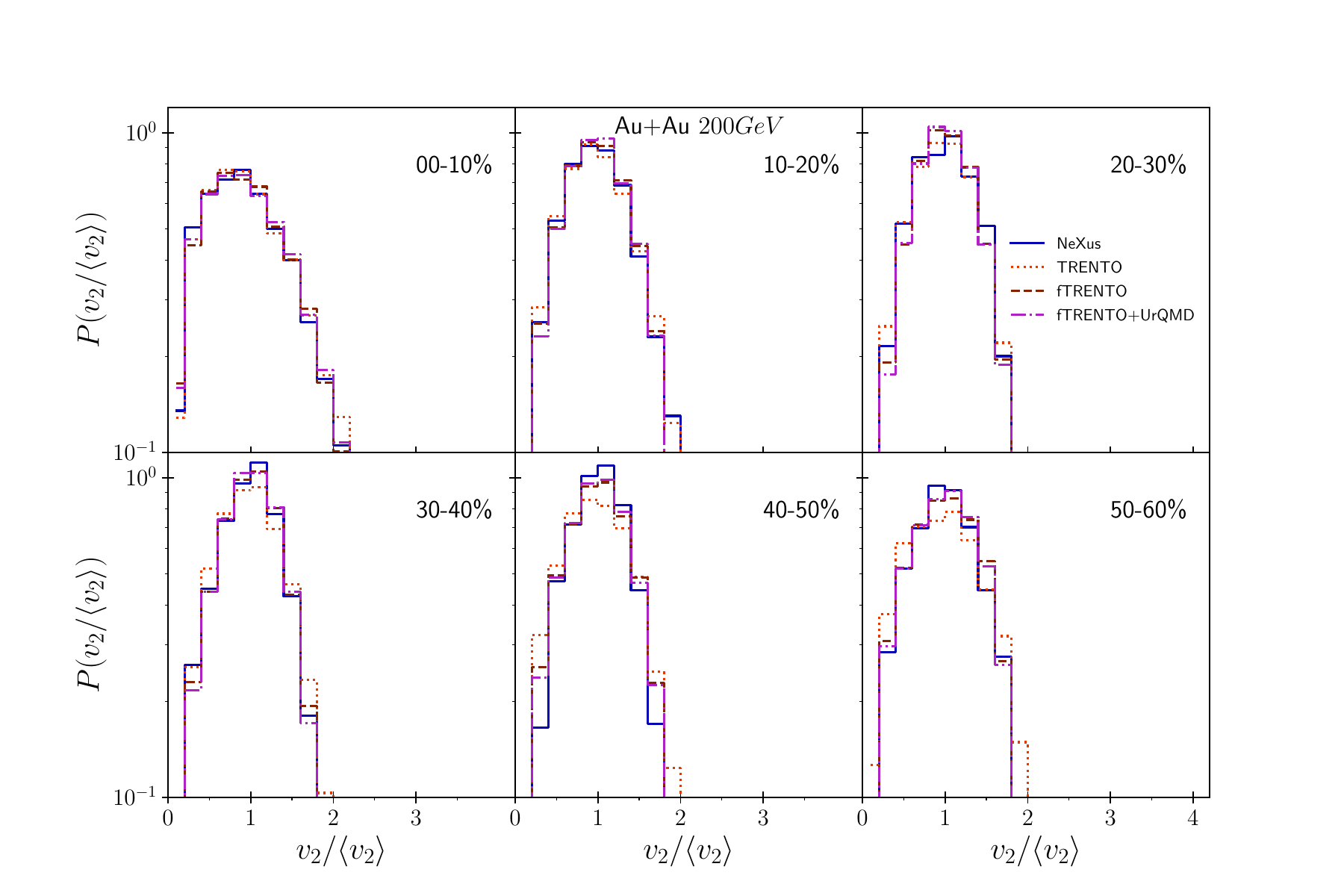}\\
      \includegraphics[width=0.8\textwidth]{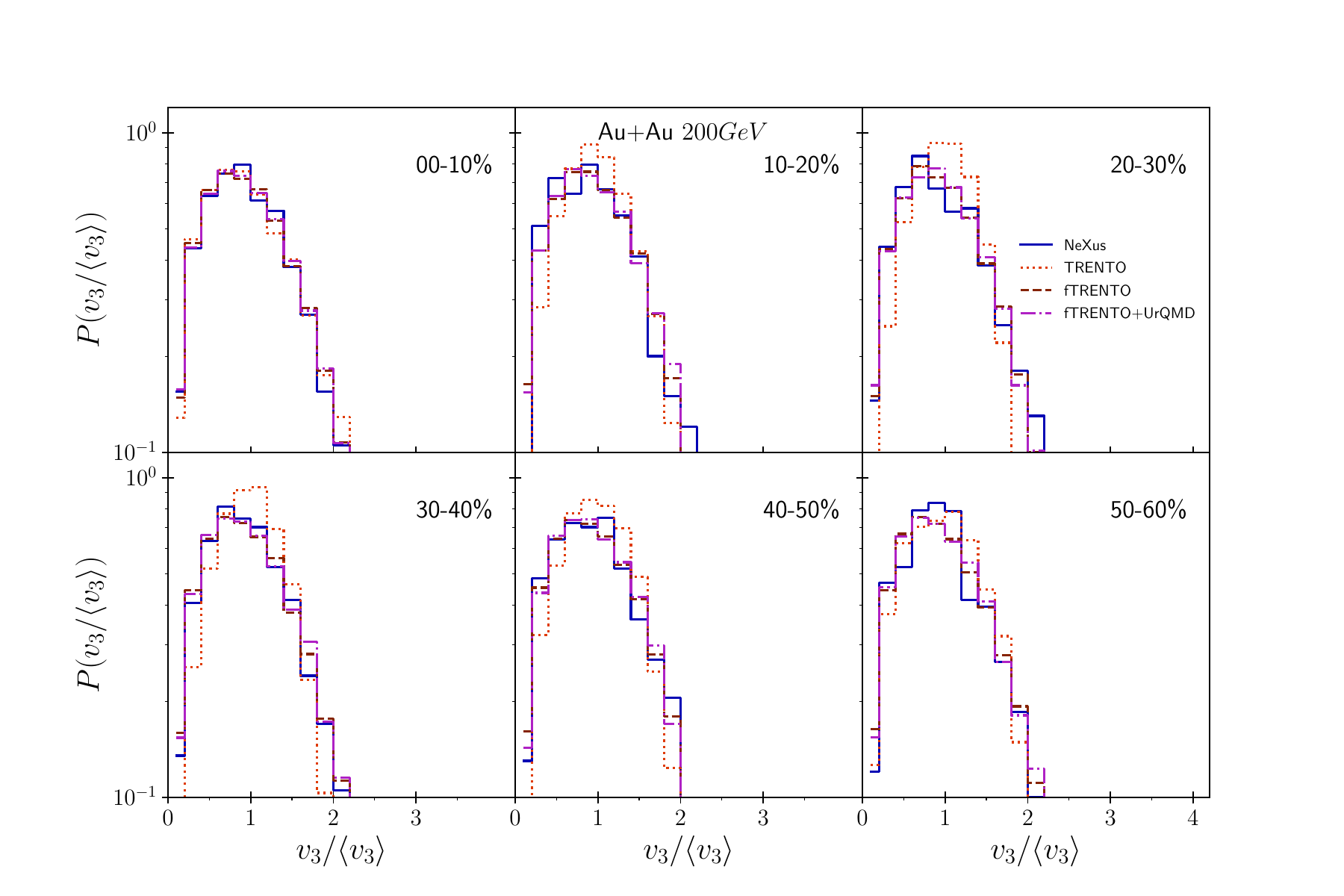}
     \caption{Top: Predictions for scaled $v_2$ distributions for NeXus and TRENTO initial conditions for Au+Au collisions at 200 GeV. For TRENTO, results of simulations with free streaming and free streaming+UrQMD are also shown.\\
Bottom: The same for n=3.     }
\label{fig:dist2}

\end{figure*}

Finally, predictions for scaled $v_n$ distributions in Au+Au collisions at 200 GeV at various centralities are shown in fig. \ref{fig:dist2}, % and \ref{fig:dist3},
comparing TRENTO and NeXus initial conditions.  Unsurprisingly, they are relatively equivalent but we do find that there are some subtle differences in their centrality dependence. 
For
$v_2$ in peripheral collisions,  NeXus  has slightly narrower fluctuations than TRENTO.  
fTRENTO and fTRENTO+UrQMD are very close to NeXus.

We now turn to the flow factorization ratios.

\begin{figure*}[!ht]

  \includegraphics[width=\linewidth]{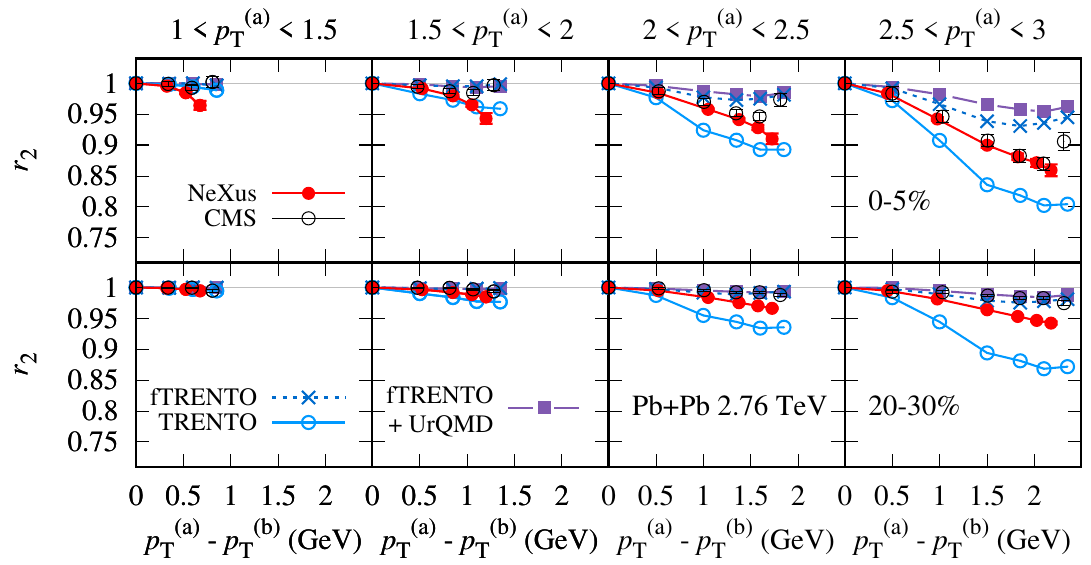}\\
  \includegraphics[width=\linewidth]{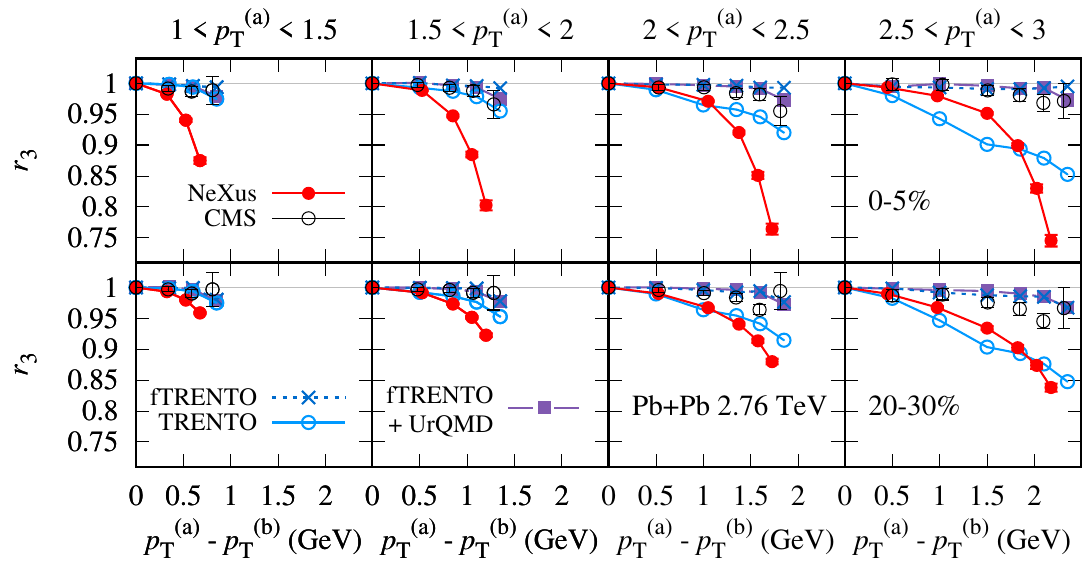}
 \caption{Results for $r_n$  for Pb+Pb collisions at 2.76 TeV:  NeXSPheRIO ($|\eta|<2.5$) (``NeXus'') and TRENTO, fTRENTO, fTRENTO+UrQMD and
   CMS data \cite{Khachatryan:2015oea} ($|\eta|<2.4$).}
\label{fig:rnLHC}
\end{figure*}

As mentioned in Sec.\ \ref{sec:rfac}, compared to CMS data \cite{Khachatryan:2015oea},  IP-Glasma+MUSIC+UrQMD  \cite{McDonald:2016vlt},  AMPT+VISHNU  \cite{Zhao:2017yhj},
TRENTO+VISHNU  \cite{Zhao:2017yhj}
provide reasonable results for $r_2$ in all centralities but $r_3$ exhibits a too large  drop for all trigger $p_T$'s  and centralities. In  fig. \ref{fig:rnLHC}, it is seen that a similar conclusion holds for    NeXSPHeRIO.
For $r_2$,
TRENTO is below data while  fTRENTO and fTRENTO+UrQMD are above.
For $r_3$,
TRENTO is again below data while  fTRENTO and fTRENTO+UrQMD are in reasonable agreement, with some overshooting.
Comparison of TRENTO (TRENTO+v-USPhydro) with \cite{Zhao:2017yhj} 
(TRENTO+VISH+UrQMD) confirms \cite{McDonald:2016vlt}
that the hadronic cascade pushes all $r_n$s towards 1 (not shown).
Comparison of fTRENTO and fTRENTO+UrQMD with TRENTO
shows a new characteristic: 
 free streaming  has a strong effect, already pushing all $r_n$s closer to 1 and leaving little space for effects from UrQMD.
  This suggests that a better treatment of the pre-hydrodynamic phase could lead to overall agreement with the $r_n$ data.

\begin{figure*}[!ht]
  \includegraphics[width=\linewidth]{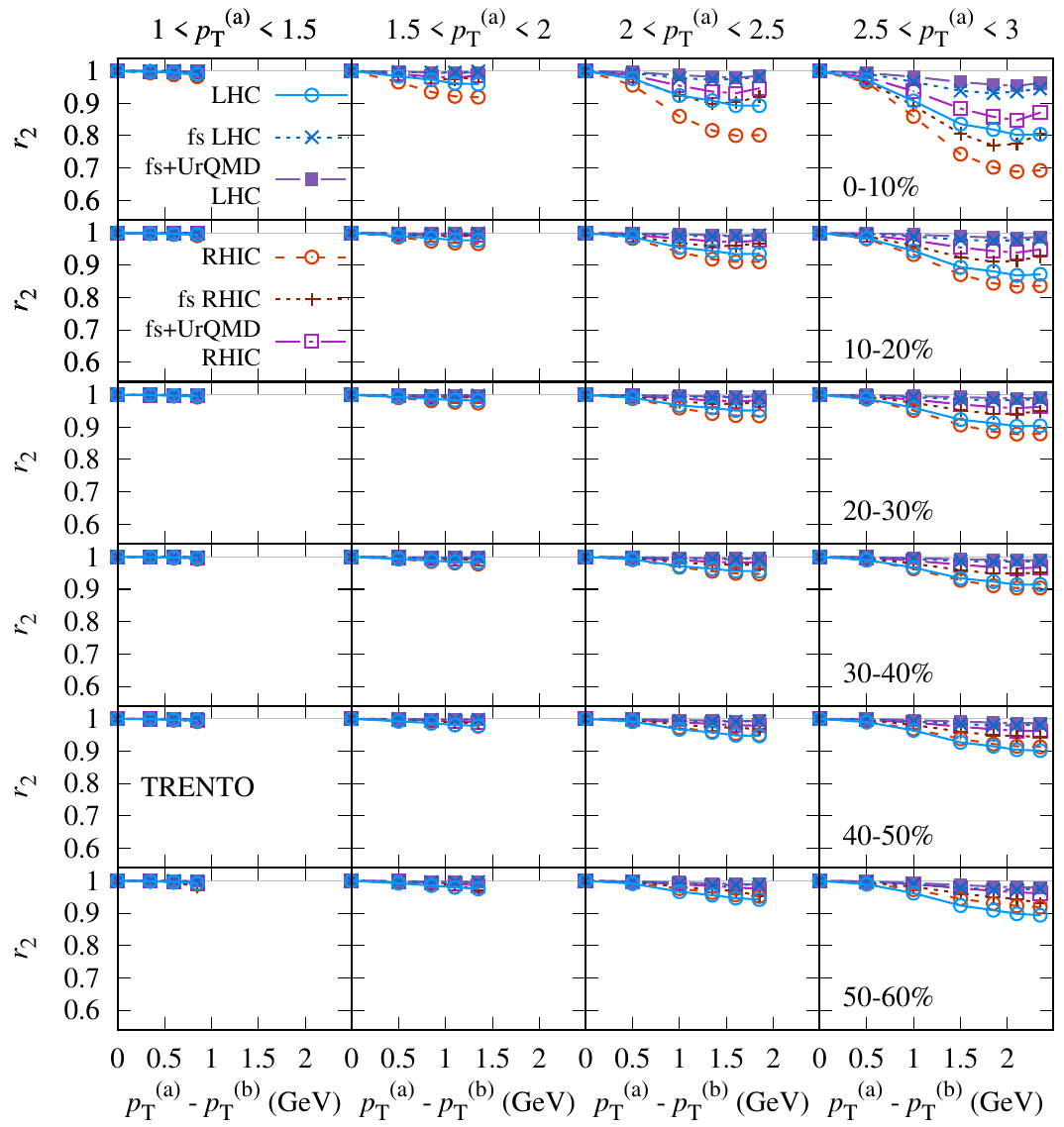}
\caption{  Comparison of  $r_2$  for TRENTO initial conditions at LHC and RHIC energies
  for various centrality windows. In this figure, LHC means TRENTO+v-USPhydro
  at 2.76 TeV, fs LHC is fTRENTO (free streaming and hydro with no UrQMD) at LHC and fs+UrQMD is fTRENTO+UrQMD (free streaming and hydro with UrQMD) at LHC. Similarly for RHIC but at  200 GeV.}
\label{fig:r2TrentoLHCRHIC}

\end{figure*}

\begin{figure*}[!ht]

  \includegraphics[width=\linewidth]{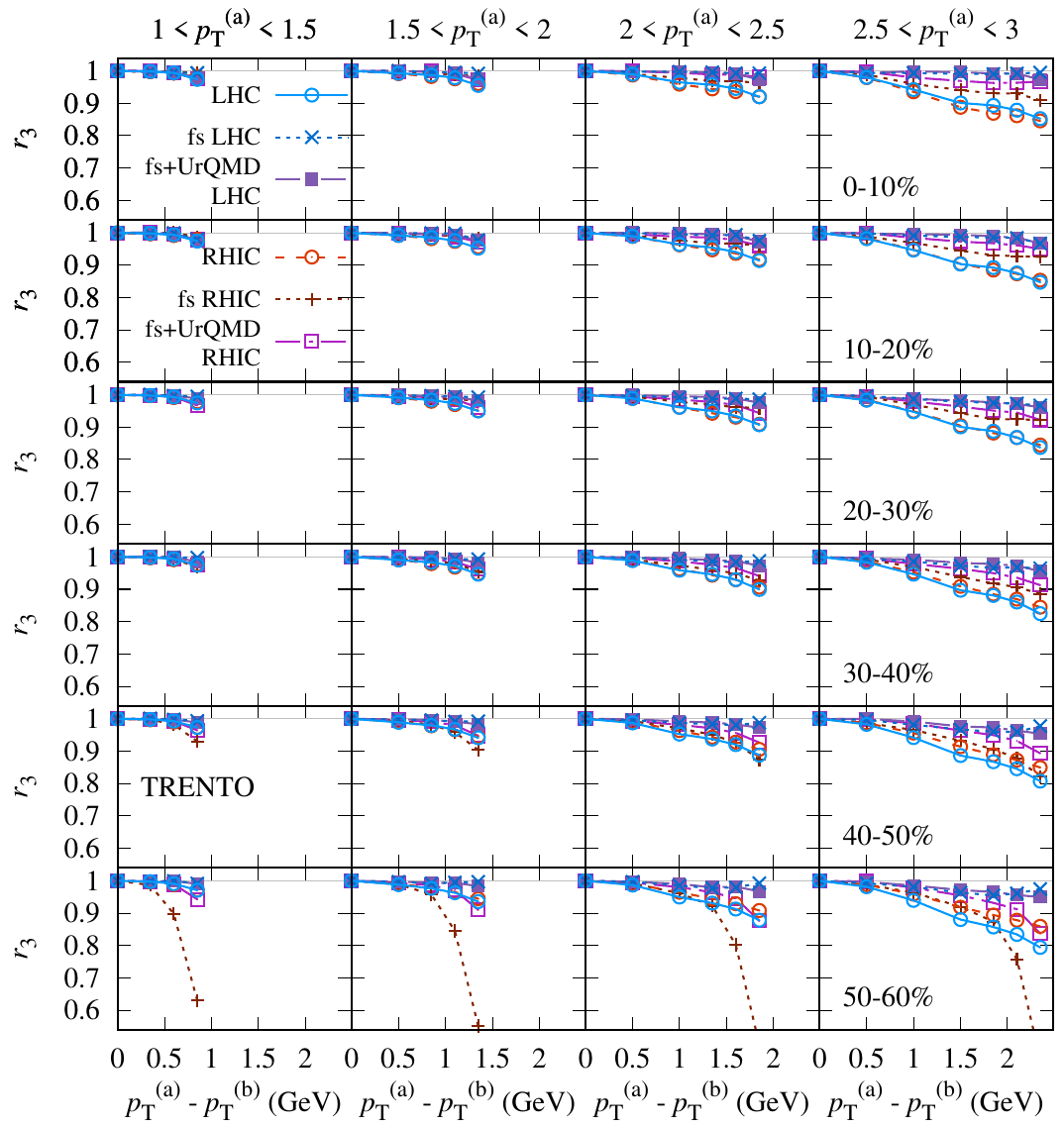}
\caption{Same as fig.\ref{fig:r2TrentoLHCRHIC} but for $r_3$.}
\label{fig:r3TrentoLHCRHIC}
\end{figure*}

We can address the energy dependence of these ratios.
Predictions for $r_n$  at PbPb 5.02  TeV as well as data and comparison at 2.76  TeV were presented 
for IP-Glasma+MUSIC+UrQMD \cite{McDonald:2016vlt}, AMPT and TRENTO \cite{Zhao:2017yhj}, both with VISHNU.
Some  difference was observed between these two LHC energies with a tendency towards more factorization breaking at lower energy.
For LHC and RHIC energies, a
comparison of factorization breaking
with TRENTO initial conditions is shown in Figs. \ref{fig:r2TrentoLHCRHIC} and \ref{fig:r3TrentoLHCRHIC}.
Factorization breaking is always larger at RHIC than at LHC, for $r_2$ (less for $r_3$) up to 40\% centrality. For less central collisions,  it becomes smaller than at LHC. Generally, central collisions at higher $p_T$ appear to have the largest factorization breaking.  However, we note that this is precisely the regime where our predictions struggle to reproduce experimental results so this regime still requires more theoretical effort. 

We note also that $r_3$ has a much weaker centrality dependence compared to $r_2$, which is likely due to the fact that $v_2$ also has a geometrical component to it when one varies centrality whereas $v_3$ is entirely fluctuations driven.  Because $v_3$ is primarily driven by fluctuations, which exist regardless of the centrality class then it is not surprising that there is not a strong centrality dependence. 

Finally, we observe that at the LHC, fTRENTO and fTRENTO+UrQMD lead to almost no factorization breaking ($r_n=1$). On the other side at RHIC, for larger $p_T$ and more central collisions, fTRENTO and fTRENTO+UrQMD $r_n$ results   are different from 1 and can be well distinguished. Therefore, given the weak dependence on viscosity, data on factorization breaking at RHIC would be useful to constrain pre-hydrodynamic  stage and final hadronic cascade.
For $r_3$ in the less central bins,  free streaming surprisingly could increase factorization breaking, but this might be due to the small differences in implementation between the initial conditions in TRENTO (with freeze out) and  fTRENTO (see Sec. \ref{sec:IC}).

\begin{figure*}[!ht]

\includegraphics[width=\linewidth]{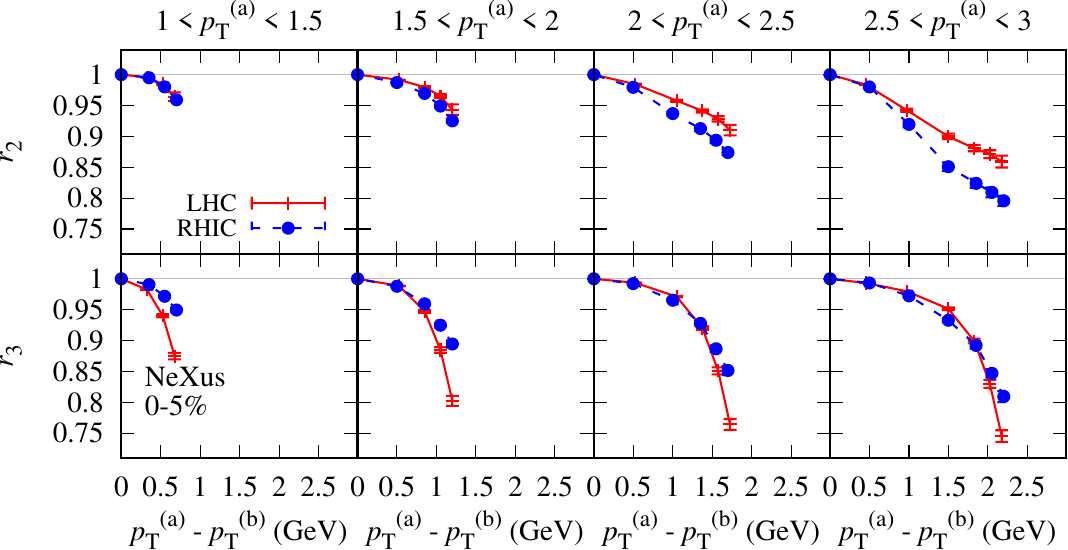}
\caption{Comparison of  $r_n$  for NeXus ($|\eta|<2.5$) at LHC and RHIC energies
  for the 0-5\% centrality window.}
\label{fig:nexusLHCRHIC}
\end{figure*}

Next we compare central NeXus initial conditions between RHIC and the LHC. A larger breaking  is seen at RHIC than at LHC  for $r_2$ (and not for $r_3$) in Fig. \ref{fig:nexusLHCRHIC}, which confirms the trend found for the TRENTO results in Figs. \ref{fig:r2TrentoLHCRHIC}-\ref{fig:r3TrentoLHCRHIC}.

\begin{figure*}[!ht]

  \includegraphics[width=\textwidth]{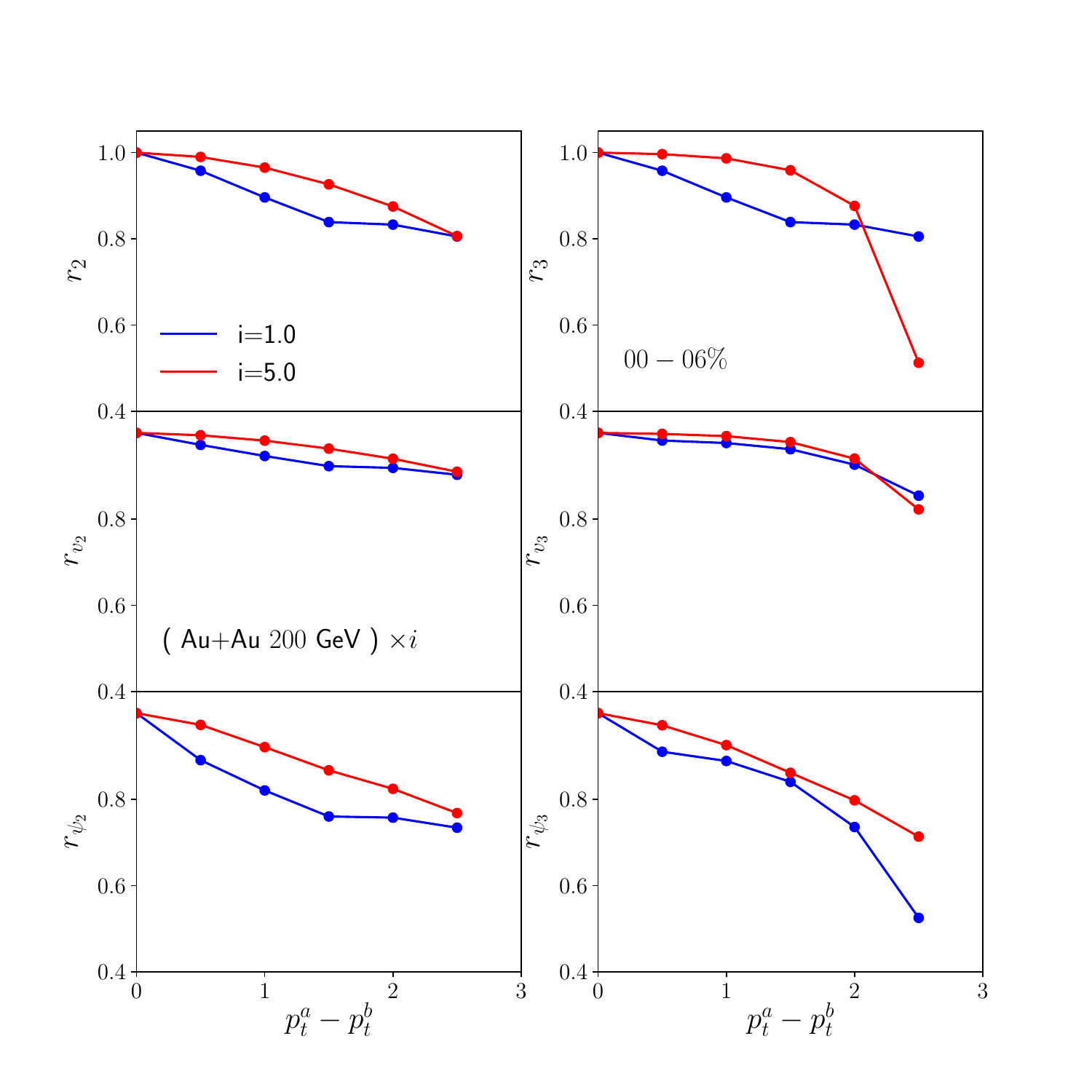}

 \caption{Comparison of factorization breaking for initial conditions at RHIC
top energy
   and the same initial conditions with energy density multiplied by a factor 5. Top
   row: $r_2$ and $r_3$, middle row:$r_{v2}$ and $r_{v3}$, bottom row:
$r_{\psi 2}$ and $r_{\psi 3}$ (definitions in text). $p_T^{(a)}=2.5\,GeV$.}
\label{fig:rescaling}
\end{figure*}

In order to understand how higher energies imply smaller breaking of factorization, we have selected 600 events in a central window and solved the ideal hydrodynamic equations at RHIC. Then exactly the same events had their energy density scaled by a factor of 5, and the hydrodynamic equations were solved. The results shown  in Fig. \ref{fig:rescaling}, indicate 
that the factorization breaking for $n=2$ (and less for $n=3$) is stronger at RHIC and this appears to be directly connected to the overall energy density scale, which leads to a smaller lifetime at RHIC vs. the LHC. Thus, one can argue that at high-energies due to the longer lifetime of hydrodynamics, one would expect $r_n\rightarrow 1$. 

To see which part of the flow vector was more affected, the following quantities (for $p_T^{(a)}=2.5\,GeV$) were computed: 
\begin{equation}\label{eqn:rvn}
    r_{vn}=\langle v_n( p_T^{(a)}) v_n( p_T^{(b)})\rangle/\sqrt{\langle v_n^2( p_T^{(a)})\rangle \langle v_n^2( p_T^{(b)})\rangle}
\end{equation}
and 
\begin{equation}\label{eqn:psi}
    r_{\psi n}=\langle \cos n(\psi_n(p_T^{(a)})-\psi_n(p_T^{(b)})) \rangle.
\end{equation}
Eq.\ (\ref{eqn:rvn}) is a Pearson coefficient between different $p_T$ bins of the flow harmonics and Eq.\ \ref{eqn:psi} demonstrates the decorrelation of event plane angles across $p_T$.  Note similar studies have been performed in \cite{Betz:2016ayq} to study high $p_T$ flow harmonics.

 Although these quantities are not independent and their product is not exactly $ r_n $, we expect that their individual behaviors can help to understand the physical picture. In Fig. \ref{fig:rescaling}, the   $r_{v_n}$ curves
are seen to be
almost energy independent, while the  $r_{\psi_n}$ ones
are energy dependent. The energy independence of $r_{v_n}$ implies that the magnitude of the $v_n$'s vs. $p_T$ generally scales in the same way regardless of the beam energy.  
The $r_{\psi_n}$ can be understood, since longer lifetimes (LHC energy) lead to different event planes tending to have the same direction due to pressure gradients in
the hydrodynamic expansion. The longer that one runs hydrodynamics, the more the event plane angles are aligned.

\begin{figure*}[!ht]

  \includegraphics[width=\linewidth]{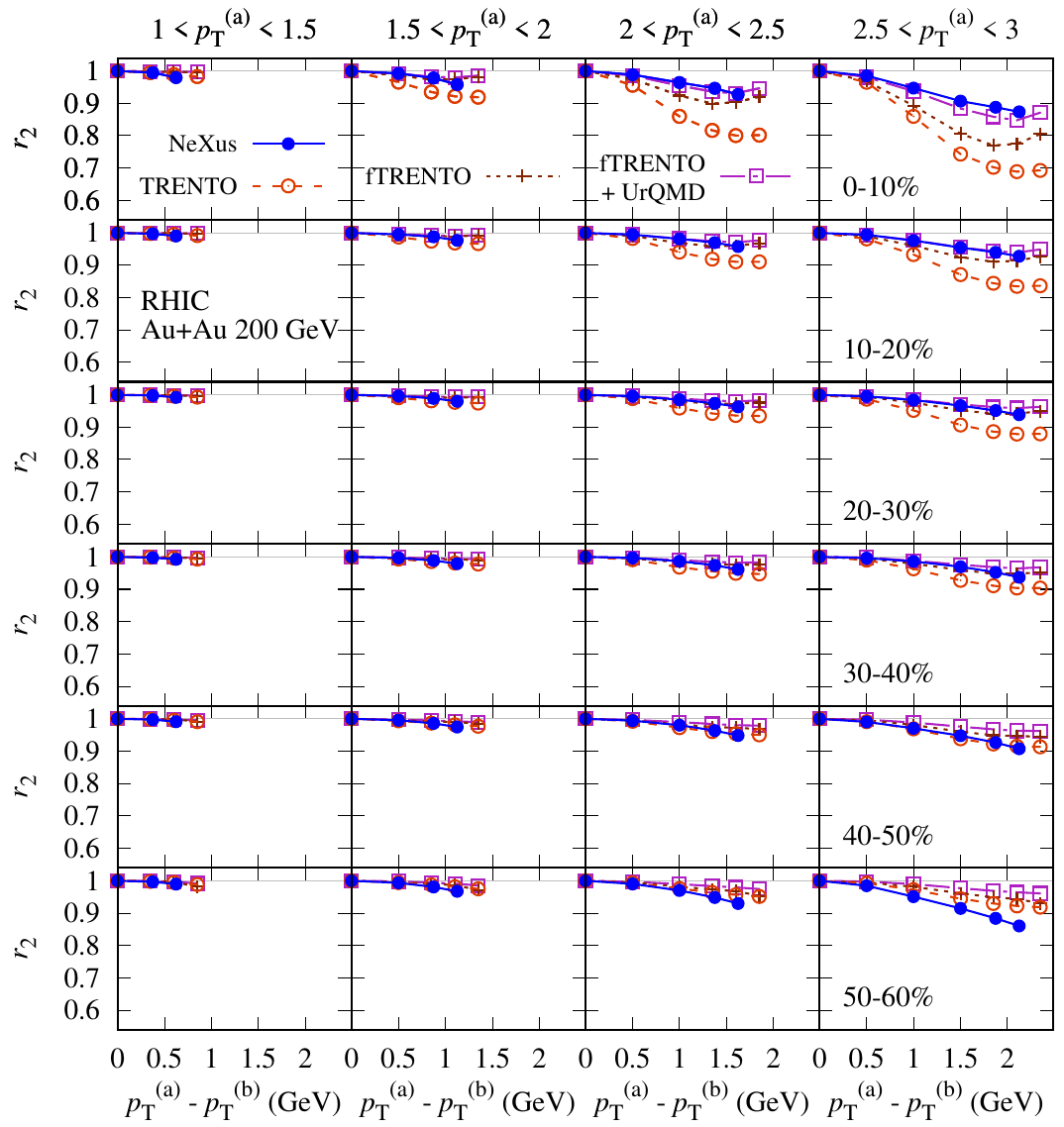}
\caption{Predictions for $r_2$ for  NeXus and TRENTO models for Au+Au collisions at 200 A GeV.}
\label{fig:r2RHIC}

\end{figure*}

\begin{figure*}[!ht]

   \includegraphics[width=\linewidth]{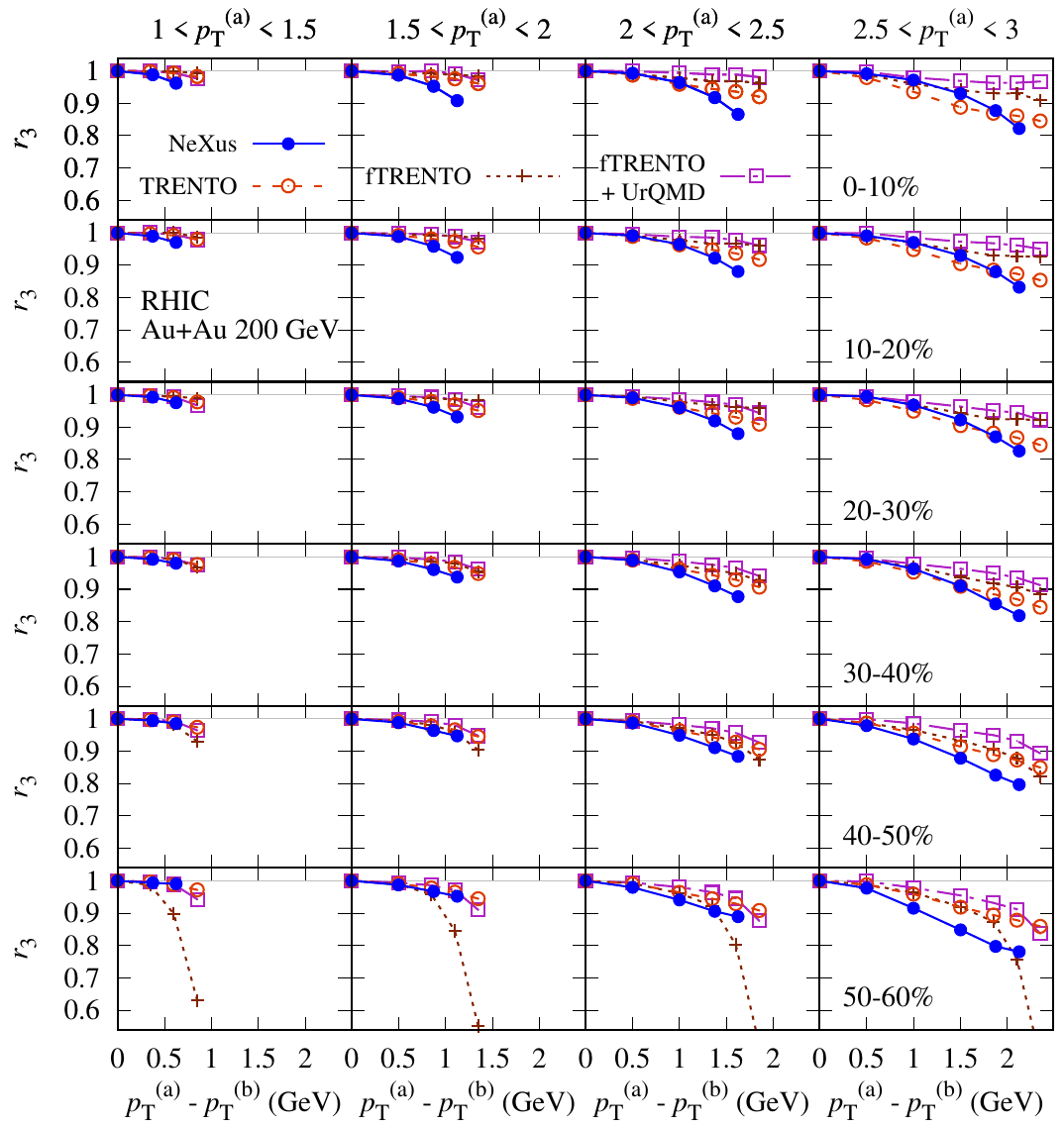}
\caption{Predictions for $r_3$ for NeXus and TRENTO models for Au+Au collisions at 200 A GeV.}
\label{fig:r3RHIC}
\end{figure*}

Now that we have better understood the scaling of the factorization with beam energy and subtle differences between Trento and NeXus initial conditions, we present results for the factorization breaking ratios at RHIC for TRENTO and NeXus 
in Figs. \ref{fig:r2RHIC}-\ref{fig:r3RHIC}. We find that for $r_2$ low $p_T$ values are nearly identical, it is only for $p_T^{(a)}$ in bins above $p_T^{(a)}> 2$ GeV that one can distinguish between TRENTO and NeXus.  Additionally, we find that high $p_T^{(a)}$ in central collisions is by far the most important region to distinguish between different types of initial conditions and the difference can be as large as $\sim$ 25 \% for 0-10\% centrality and larger for a more central window. For comparison, at LHC, for $r_2$ in Fig.  \ref{fig:r2TrentoLHCRHIC} (in the 0-5\% bin), differences between NeXus and TRENTO were smaller than 10\%. Thus, it would be interesting to have factorization
breaking measurements from RHIC to further distinguish between initial state models.

Comparing our results for both $r_2$ ad $r_3$ we find that $r_2$ also is a better candidate for distinguishing initial state models.  While subtle differences exist for $r_3$ for TRENTO vs. NeXus in Fig. \ref{fig:r3RHIC}, they would require much more precise experimental data to distinguish between models (and as already mentioned, the theoretical understanding of $r_3$ is not yet completely clear).

\section{Conclusion}

 Because RHIC also is more sensitive to a variety of medium effects compared to the LHC (i.e. the finite $\mu_B$ equation of state, diffusion, criticality etc), it is that much more important to find medium independent observables that can help to constrain initial state effects. In this paper we  concentrate on two quantities that are not very sensitive to viscosity and compare predictions from two initial condition models, NeXus and TRENTO (with $p=0$ configuration). We found that 
the scaled $v_n$ distributions that were used to rule out initial condition models at the LHC do not provide new constraints at top RHIC energy. Here we argue that factorization breaking will be sensitive to the choice in the initial state at RHIC and outline the best centrality and $p_T$ windows to study.  Generally, we find that $r_2$ is a better candidate for distinguishing initial state models and specifically in central collisions and using higher $p_T$ cuts.

Our main conclusion is that factorization breaking ratios (as function of $p_T$)
would be very interesting to get experimentally at top RHIC energy because
1) they should be an excellent tool to constrain initial condition models (as illustrated here for NeXus and TRENTO), differences in  $r_n$ being enhanced at RHIC compared to LHC,
2) they should also allow to put constrain on early free streaming and late hadronic transport (shown here for fTRENTO and fTRENTO+UrQMD),
again because differences appear more clearly at RHIC than at the LHC (we expect that free streaming and UrQMD would act with the same order of magnitude independently of the initial condition model).

In order to understand in more detail how the factorization breaking observable scales with beam energy,
in particular while it has stronger effects at RHIC than LHC,
we have calculated two quantities that correlate either the magnitude of flow harmonics across $p_T$, $r_{vn}$, or the correlation between event-plane angles across $p_T$, $r_{\psi n}$.  We found that the correlation between overall magnitudes of flow harmonics is not strongly sensitive to the lifetime of hydrodynamics.  In contrast, though, the correlation of event plan angles is more sensitive to the lifetime of the system. 
 
 While not done in this initial study, it may be interesting to also calculate the sensitivity of factorization breaking ratios at RHIC for small systems and lower beam energies. We leave this as a potential future work.

\section{Acknowledgements}
L.B.~thanks support from Conselho Nacional de Desenvolvimento Cient\'{\i}fico e Tecnol\'ogico (CNPq).
F.G.G. was supported by Conselho Nacional de Desenvolvimento Cient\'{\i}fico  e  Tecnol\'ogico  (CNPq grant 312203/2015-2)  and FAPEMIG (grant APQ-02107-16).
F.G.~acknowledges  support  from
Funda\c{c}\~ao de Amparo \`a Pesquisa do Estado de S\~ao Paulo
(FAPESP  grant  2018/24720-6)%, Conselho Nacional de Desenvolvimento Cient\'{\i}fico e Tecnol\'ogico (CNPq grant 310141/2016-8)
and project INCT-FNA Proc.~No.~464898/2014-5.
J.N.H. acknowledges the support from the US-DOE Nuclear Science Grant No. DE-SC0020633. M.H. was supported in part by the National Science Foundation (NSF) within the framework of the MUSES collaboration, under grant number OAC-2103680. 
P.I.~thanks support from Coordena\c{c}\~ao de Aperfei\c{c}oamento de Pessoal de N\'{\i}vel Superior (CAPES) and
Conselho Nacional de Desenvolvimento Cient\'{\i}fico e Tecnol\'ogico (CNPq grant 142151/2019-0).
M.L.~acknowledges support from FAPESP projects 2016/24029-6  and 2017/05685-2, and project INCT-FNA Proc.~No.~464898/2014-5.

\bibliography{bibli}

\providecommand{\noopsort}[1]{}\providecommand{\singleletter}[1]{#1}%
\begin{thebibliography}{89}%
\makeatletter
\providecommand \@ifxundefined [1]{%
 \@ifx{#1\undefined}
}%
\providecommand \@ifnum [1]{%
 \ifnum #1\expandafter \@firstoftwo
 \else \expandafter \@secondoftwo
 \fi
}%
\providecommand \@ifx [1]{%
 \ifx #1\expandafter \@firstoftwo
 \else \expandafter \@secondoftwo
 \fi
}%
\providecommand \natexlab [1]{#1}%
\providecommand \enquote  [1]{``#1''}%
\providecommand \bibnamefont  [1]{#1}%
\providecommand \bibfnamefont [1]{#1}%
\providecommand \citenamefont [1]{#1}%
\providecommand \href@noop [0]{\@secondoftwo}%
\providecommand \href [0]{\begingroup \@sanitize@url \@href}%
\providecommand \@href[1]{\@@startlink{#1}\@@href}%
\providecommand \@@href[1]{\endgroup#1\@@endlink}%
\providecommand \@sanitize@url [0]{\catcode `\\12\catcode `\$12\catcode
  `\&12\catcode `\#12\catcode `\^12\catcode `\_12\catcode `\%12\relax}%
\providecommand \@@startlink[1]{}%
\providecommand \@@endlink[0]{}%
\providecommand \url  [0]{\begingroup\@sanitize@url \@url }%
\providecommand \@url [1]{\endgroup\@href {#1}{\urlprefix }}%
\providecommand \urlprefix  [0]{URL }%
\providecommand \Eprint [0]{\href }%
\providecommand \doibase [0]{http://dx.doi.org/}%
\providecommand \selectlanguage [0]{\@gobble}%
\providecommand \bibinfo  [0]{\@secondoftwo}%
\providecommand \bibfield  [0]{\@secondoftwo}%
\providecommand \translation [1]{[#1]}%
\providecommand \BibitemOpen [0]{}%
\providecommand \bibitemStop [0]{}%
\providecommand \bibitemNoStop [0]{.\EOS\space}%
\providecommand \EOS [0]{\spacefactor3000\relax}%
\providecommand \BibitemShut  [1]{\csname bibitem#1\endcsname}%
\let\auto@bib@innerbib\@empty
%</preamble>
\bibitem [{\citenamefont {Ratti}(2018)}]{Ratti:2018ksb}%
  \BibitemOpen
  \bibfield  {author} {\bibinfo {author} {\bibfnamefont {C.}~\bibnamefont
  {Ratti}},\ }\href {\doibase 10.1088/1361-6633/aabb97} {\bibfield  {journal}
  {\bibinfo  {journal} {Rept. Prog. Phys.}\ }\textbf {\bibinfo {volume} {81}},\
  \bibinfo {pages} {084301} (\bibinfo {year} {2018})},\ \Eprint
  {http://arxiv.org/abs/1804.07810} {arXiv:1804.07810 [hep-lat]} \BibitemShut
  {NoStop}%
\bibitem [{\citenamefont {Bzdak}\ \emph {et~al.}(2020)\citenamefont {Bzdak},
  \citenamefont {Esumi}, \citenamefont {Koch}, \citenamefont {Liao},
  \citenamefont {Stephanov},\ and\ \citenamefont {Xu}}]{Bzdak:2019pkr}%
  \BibitemOpen
  \bibfield  {author} {\bibinfo {author} {\bibfnamefont {A.}~\bibnamefont
  {Bzdak}}, \bibinfo {author} {\bibfnamefont {S.}~\bibnamefont {Esumi}},
  \bibinfo {author} {\bibfnamefont {V.}~\bibnamefont {Koch}}, \bibinfo {author}
  {\bibfnamefont {J.}~\bibnamefont {Liao}}, \bibinfo {author} {\bibfnamefont
  {M.}~\bibnamefont {Stephanov}}, \ and\ \bibinfo {author} {\bibfnamefont
  {N.}~\bibnamefont {Xu}},\ }\href {\doibase 10.1016/j.physrep.2020.01.005}
  {\bibfield  {journal} {\bibinfo  {journal} {Phys. Rept.}\ }\textbf {\bibinfo
  {volume} {853}},\ \bibinfo {pages} {1} (\bibinfo {year} {2020})},\ \Eprint
  {http://arxiv.org/abs/1906.00936} {arXiv:1906.00936 [nucl-th]} \BibitemShut
  {NoStop}%
\bibitem [{\citenamefont {Dexheimer}\ \emph {et~al.}(2020)\citenamefont
  {Dexheimer}, \citenamefont {Noronha}, \citenamefont {Noronha-Hostler},
  \citenamefont {Ratti},\ and\ \citenamefont {Yunes}}]{Dexheimer:2020zzs}%
  \BibitemOpen
  \bibfield  {author} {\bibinfo {author} {\bibfnamefont {V.}~\bibnamefont
  {Dexheimer}}, \bibinfo {author} {\bibfnamefont {J.}~\bibnamefont {Noronha}},
  \bibinfo {author} {\bibfnamefont {J.}~\bibnamefont {Noronha-Hostler}},
  \bibinfo {author} {\bibfnamefont {C.}~\bibnamefont {Ratti}}, \ and\ \bibinfo
  {author} {\bibfnamefont {N.}~\bibnamefont {Yunes}},\ }\href@noop {} {\
  (\bibinfo {year} {2020})},\ \Eprint {http://arxiv.org/abs/2010.08834}
  {arXiv:2010.08834 [nucl-th]} \BibitemShut {NoStop}%
\bibitem [{\citenamefont {Monnai}\ \emph {et~al.}(2021)\citenamefont {Monnai},
  \citenamefont {Schenke},\ and\ \citenamefont {Shen}}]{Monnai:2021kgu}%
  \BibitemOpen
  \bibfield  {author} {\bibinfo {author} {\bibfnamefont {A.}~\bibnamefont
  {Monnai}}, \bibinfo {author} {\bibfnamefont {B.}~\bibnamefont {Schenke}}, \
  and\ \bibinfo {author} {\bibfnamefont {C.}~\bibnamefont {Shen}},\ }\href
  {\doibase 10.1142/S0217751X21300076} {\bibfield  {journal} {\bibinfo
  {journal} {Int. J. Mod. Phys. A}\ }\textbf {\bibinfo {volume} {36}},\
  \bibinfo {pages} {2130007} (\bibinfo {year} {2021})},\ \Eprint
  {http://arxiv.org/abs/2101.11591} {arXiv:2101.11591 [nucl-th]} \BibitemShut
  {NoStop}%
\bibitem [{\citenamefont {STARcollaboration}(2014)}]{STARnote}%
  \BibitemOpen
  \bibfield  {author} {\bibinfo {author} {\bibnamefont {STARcollaboration}},\
  }\href@noop {} {\enquote {\bibinfo {title} {{Studying the Phase Diagram of
  QCD Matter at RHIC}},}\ } (\bibinfo {year} {2014})\BibitemShut {NoStop}%
\bibitem [{\citenamefont {Cebra}\ \emph {et~al.}(2014)\citenamefont {Cebra},
  \citenamefont {Brovko}, \citenamefont {Flores}, \citenamefont {Haag},\ and\
  \citenamefont {Klay}}]{Cebra:2014sxa}%
  \BibitemOpen
  \bibfield  {author} {\bibinfo {author} {\bibfnamefont {D.}~\bibnamefont
  {Cebra}}, \bibinfo {author} {\bibfnamefont {S.~G.}\ \bibnamefont {Brovko}},
  \bibinfo {author} {\bibfnamefont {C.~E.}\ \bibnamefont {Flores}}, \bibinfo
  {author} {\bibfnamefont {B.~A.}\ \bibnamefont {Haag}}, \ and\ \bibinfo
  {author} {\bibfnamefont {J.~L.}\ \bibnamefont {Klay}},\ }\href@noop {} {\
  (\bibinfo {year} {2014})},\ \Eprint {http://arxiv.org/abs/1408.1369}
  {arXiv:1408.1369 [nucl-ex]} \BibitemShut {NoStop}%
\bibitem [{\citenamefont {Galatyuk}(2014)}]{Galatyuk:2014vha}%
  \BibitemOpen
  \bibfield  {author} {\bibinfo {author} {\bibfnamefont {T.}~\bibnamefont
  {Galatyuk}} (\bibinfo {collaboration} {HADES}),\ }\href {\doibase
  10.1016/j.nuclphysa.2014.10.044} {\bibfield  {journal} {\bibinfo  {journal}
  {Nucl. Phys. A}\ }\textbf {\bibinfo {volume} {931}},\ \bibinfo {pages} {41}
  (\bibinfo {year} {2014})}\BibitemShut {NoStop}%
\bibitem [{\citenamefont {Friese}(2006)}]{Friese:2006dj}%
  \BibitemOpen
  \bibfield  {author} {\bibinfo {author} {\bibfnamefont {V.}~\bibnamefont
  {Friese}},\ }\href {\doibase 10.1016/j.nuclphysa.2006.06.018} {\bibfield
  {journal} {\bibinfo  {journal} {Nucl. Phys. A}\ }\textbf {\bibinfo {volume}
  {774}},\ \bibinfo {pages} {377} (\bibinfo {year} {2006})}\BibitemShut
  {NoStop}%
\bibitem [{\citenamefont {Tahir}\ \emph {et~al.}(2005)\citenamefont {Tahir}
  \emph {et~al.}}]{Tahir:2005zz}%
  \BibitemOpen
  \bibfield  {author} {\bibinfo {author} {\bibfnamefont {N.~A.}\ \bibnamefont
  {Tahir}} \emph {et~al.},\ }\href {\doibase 10.1103/PhysRevLett.95.035001}
  {\bibfield  {journal} {\bibinfo  {journal} {Phys. Rev. Lett.}\ }\textbf
  {\bibinfo {volume} {95}},\ \bibinfo {pages} {035001} (\bibinfo {year}
  {2005})}\BibitemShut {NoStop}%
\bibitem [{\citenamefont {Lutz}\ \emph {et~al.}(2009)\citenamefont {Lutz} \emph
  {et~al.}}]{Lutz:2009ff}%
  \BibitemOpen
  \bibfield  {author} {\bibinfo {author} {\bibfnamefont {M.~F.~M.}\
  \bibnamefont {Lutz}} \emph {et~al.} (\bibinfo {collaboration} {PANDA}),\
  }\href@noop {} {\  (\bibinfo {year} {2009})},\ \Eprint
  {http://arxiv.org/abs/0903.3905} {arXiv:0903.3905 [hep-ex]} \BibitemShut
  {NoStop}%
\bibitem [{\citenamefont {Durante}\ \emph {et~al.}(2019)\citenamefont {Durante}
  \emph {et~al.}}]{Durante:2019hzd}%
  \BibitemOpen
  \bibfield  {author} {\bibinfo {author} {\bibfnamefont {M.}~\bibnamefont
  {Durante}} \emph {et~al.},\ }\href {\doibase 10.1088/1402-4896/aaf93f}
  {\bibfield  {journal} {\bibinfo  {journal} {Phys. Scripta}\ }\textbf
  {\bibinfo {volume} {94}},\ \bibinfo {pages} {033001} (\bibinfo {year}
  {2019})},\ \Eprint {http://arxiv.org/abs/1903.05693} {arXiv:1903.05693
  [nucl-th]} \BibitemShut {NoStop}%
\bibitem [{\citenamefont {Kekelidze}\ \emph {et~al.}(2017)\citenamefont
  {Kekelidze}, \citenamefont {Kovalenko}, \citenamefont {Lednicky},
  \citenamefont {Matveev}, \citenamefont {Meshkov}, \citenamefont {Sorin},\
  and\ \citenamefont {Trubnikov}}]{Kekelidze:2017tgp}%
  \BibitemOpen
  \bibfield  {author} {\bibinfo {author} {\bibfnamefont {V.}~\bibnamefont
  {Kekelidze}}, \bibinfo {author} {\bibfnamefont {A.}~\bibnamefont
  {Kovalenko}}, \bibinfo {author} {\bibfnamefont {R.}~\bibnamefont {Lednicky}},
  \bibinfo {author} {\bibfnamefont {V.}~\bibnamefont {Matveev}}, \bibinfo
  {author} {\bibfnamefont {I.}~\bibnamefont {Meshkov}}, \bibinfo {author}
  {\bibfnamefont {A.}~\bibnamefont {Sorin}}, \ and\ \bibinfo {author}
  {\bibfnamefont {G.}~\bibnamefont {Trubnikov}},\ }\href {\doibase
  10.1016/j.nuclphysa.2017.06.031} {\bibfield  {journal} {\bibinfo  {journal}
  {Nucl. Phys. A}\ }\textbf {\bibinfo {volume} {967}},\ \bibinfo {pages} {884}
  (\bibinfo {year} {2017})}\BibitemShut {NoStop}%
\bibitem [{\citenamefont {Kekelidze}\ \emph {et~al.}(2016)\citenamefont
  {Kekelidze}, \citenamefont {Kovalenko}, \citenamefont {Lednicky},
  \citenamefont {Matveev}, \citenamefont {Meshkov}, \citenamefont {Sorin},\
  and\ \citenamefont {Trubnikov}}]{Kekelidze:2016wkp}%
  \BibitemOpen
  \bibfield  {author} {\bibinfo {author} {\bibfnamefont {V.}~\bibnamefont
  {Kekelidze}}, \bibinfo {author} {\bibfnamefont {A.}~\bibnamefont
  {Kovalenko}}, \bibinfo {author} {\bibfnamefont {R.}~\bibnamefont {Lednicky}},
  \bibinfo {author} {\bibfnamefont {V.}~\bibnamefont {Matveev}}, \bibinfo
  {author} {\bibfnamefont {I.}~\bibnamefont {Meshkov}}, \bibinfo {author}
  {\bibfnamefont {A.}~\bibnamefont {Sorin}}, \ and\ \bibinfo {author}
  {\bibfnamefont {G.}~\bibnamefont {Trubnikov}},\ }\href {\doibase
  10.1016/j.nuclphysa.2016.03.019} {\bibfield  {journal} {\bibinfo  {journal}
  {Nucl. Phys. A}\ }\textbf {\bibinfo {volume} {956}},\ \bibinfo {pages} {846}
  (\bibinfo {year} {2016})}\BibitemShut {NoStop}%
\bibitem [{\citenamefont {Heinz}\ and\ \citenamefont
  {Snellings}(2013)}]{reviewHeinz}%
  \BibitemOpen
  \bibfield  {author} {\bibinfo {author} {\bibfnamefont {U.}~\bibnamefont
  {Heinz}}\ and\ \bibinfo {author} {\bibfnamefont {R.}~\bibnamefont
  {Snellings}},\ }\href@noop {} {\bibfield  {journal} {\bibinfo  {journal}
  {Ann. Rev. Nucl. Part. Sci.}\ }\textbf {\bibinfo {volume} {63}},\ \bibinfo
  {pages} {123} (\bibinfo {year} {2013})},\ \Eprint
  {http://arxiv.org/abs/arXiv:1301.2826} {arXiv:1301.2826} \BibitemShut
  {NoStop}%
\bibitem [{\citenamefont {Gale}\ \emph
  {et~al.}(2013{\natexlab{a}})\citenamefont {Gale}, \citenamefont {Jeon},\ and\
  \citenamefont {Schenke}}]{reviewGale}%
  \BibitemOpen
  \bibfield  {author} {\bibinfo {author} {\bibfnamefont {C.}~\bibnamefont
  {Gale}}, \bibinfo {author} {\bibfnamefont {S.}~\bibnamefont {Jeon}}, \ and\
  \bibinfo {author} {\bibfnamefont {B.}~\bibnamefont {Schenke}},\ }\href@noop
  {} {\bibfield  {journal} {\bibinfo  {journal} {Int. J. Mod. Phys. A}\
  }\textbf {\bibinfo {volume} {28}},\ \bibinfo {pages} {1340011} (\bibinfo
  {year} {2013}{\natexlab{a}})},\ \Eprint
  {http://arxiv.org/abs/arXiv:1301.5893} {arXiv:1301.5893} \BibitemShut
  {NoStop}%
\bibitem [{\citenamefont {de~Souza}\ \emph {et~al.}(2016)\citenamefont
  {de~Souza}, \citenamefont {Koide},\ and\ \citenamefont
  {Kodama}}]{reviewdeSouza}%
  \BibitemOpen
  \bibfield  {author} {\bibinfo {author} {\bibfnamefont {R.~D.}\ \bibnamefont
  {de~Souza}}, \bibinfo {author} {\bibfnamefont {T.}~\bibnamefont {Koide}}, \
  and\ \bibinfo {author} {\bibfnamefont {T.}~\bibnamefont {Kodama}},\
  }\href@noop {} {\bibfield  {journal} {\bibinfo  {journal} {Prog. Part. Nucl.
  Phys.}\ }\textbf {\bibinfo {volume} {86}},\ \bibinfo {pages} {35} (\bibinfo
  {year} {2016})},\ \Eprint {http://arxiv.org/abs/arXiv:1506.03863}
  {arXiv:1506.03863} \BibitemShut {NoStop}%
\bibitem [{\citenamefont {Jeon}\ and\ \citenamefont {Heinz}()}]{reviewQGP5}%
  \BibitemOpen
  \bibfield  {author} {\bibinfo {author} {\bibfnamefont {S.}~\bibnamefont
  {Jeon}}\ and\ \bibinfo {author} {\bibfnamefont {U.}~\bibnamefont {Heinz}},\
  }\href@noop {} {\enquote {\bibinfo {title} {Quark gluon plasma 5},}\ }\Eprint
  {http://arxiv.org/abs/arXiv:1503.03931} {arXiv:1503.03931} \BibitemShut
  {NoStop}%
\bibitem [{\citenamefont {Giacalone}\ \emph
  {et~al.}(2017{\natexlab{a}})\citenamefont {Giacalone}, \citenamefont
  {Noronha-Hostler},\ and\ \citenamefont {Ollitrault}}]{Giacalone:2017uqx}%
  \BibitemOpen
  \bibfield  {author} {\bibinfo {author} {\bibfnamefont {G.}~\bibnamefont
  {Giacalone}}, \bibinfo {author} {\bibfnamefont {J.}~\bibnamefont
  {Noronha-Hostler}}, \ and\ \bibinfo {author} {\bibfnamefont {J.-Y.}\
  \bibnamefont {Ollitrault}},\ }\href {\doibase 10.1103/PhysRevC.95.054910}
  {\bibfield  {journal} {\bibinfo  {journal} {Phys. Rev.}\ }\textbf {\bibinfo
  {volume} {C95}},\ \bibinfo {pages} {054910} (\bibinfo {year}
  {2017}{\natexlab{a}})},\ \Eprint {http://arxiv.org/abs/1702.01730}
  {arXiv:1702.01730 [nucl-th]} \BibitemShut {NoStop}%
%%CITATION = ARXIV:1702.01730;%%
\bibitem [{\citenamefont {J.Jia}(2013)}]{ATLAS12}%
  \BibitemOpen
  \bibfield  {author} {\bibinfo {author} {\bibnamefont {J.Jia}} (\bibinfo
  {collaboration} {ATLAS collaboration}),\ }\href@noop {} {\bibfield  {journal}
  {\bibinfo  {journal} {Nucl. Phys. A}\ }\textbf {\bibinfo {volume}
  {904-905}},\ \bibinfo {pages} {421c} (\bibinfo {year} {2013})},\ \Eprint
  {http://arxiv.org/abs/arXiv:1209.4232} {arXiv:1209.4232} \BibitemShut
  {NoStop}%
\bibitem [{\citenamefont {Aad}\ \emph {et~al.}(2013)\citenamefont {Aad} \emph
  {et~al.}}]{ATLAS13}%
  \BibitemOpen
  \bibfield  {author} {\bibinfo {author} {\bibfnamefont {G.}~\bibnamefont
  {Aad}} \emph {et~al.} (\bibinfo {collaboration} {ATLAS collaboration}),\
  }\href@noop {} {\bibfield  {journal} {\bibinfo  {journal} {JHEP}\ }\textbf
  {\bibinfo {volume} {1311}},\ \bibinfo {pages} {183} (\bibinfo {year}
  {2013})},\ \Eprint {http://arxiv.org/abs/arXiv:1305.2942} {arXiv:1305.2942}
  \BibitemShut {NoStop}%
\bibitem [{\citenamefont {A.R.Timmins}(2013)}]{ALICE13}%
  \BibitemOpen
  \bibfield  {author} {\bibinfo {author} {\bibnamefont {A.R.Timmins}} (\bibinfo
  {collaboration} {ALICE collaboration}),\ }\href@noop {} {\bibfield  {journal}
  {\bibinfo  {journal} {J. Phys. Conf. Ser.}\ }\textbf {\bibinfo {volume}
  {446}},\ \bibinfo {pages} {012031} (\bibinfo {year} {2013})},\ \Eprint
  {http://arxiv.org/abs/arXiv:1301.6084} {arXiv:1301.6084} \BibitemShut
  {NoStop}%
\bibitem [{\citenamefont {Khachatryan}\ \emph {et~al.}(2015)\citenamefont
  {Khachatryan} \emph {et~al.}}]{Khachatryan:2015oea}%
  \BibitemOpen
  \bibfield  {author} {\bibinfo {author} {\bibfnamefont {V.}~\bibnamefont
  {Khachatryan}} \emph {et~al.} (\bibinfo {collaboration} {CMS}),\ }\href
  {\doibase 10.1103/PhysRevC.92.034911} {\bibfield  {journal} {\bibinfo
  {journal} {Phys. Rev.}\ }\textbf {\bibinfo {volume} {C92}},\ \bibinfo {pages}
  {034911} (\bibinfo {year} {2015})},\ \Eprint
  {http://arxiv.org/abs/1503.01692} {arXiv:1503.01692 [nucl-ex]} \BibitemShut
  {NoStop}%
%%CITATION = ARXIV:1503.01692;%%
\bibitem [{\citenamefont {Acharya}\ \emph {et~al.}(2017)\citenamefont {Acharya}
  \emph {et~al.}}]{Acharya:2017ino}%
  \BibitemOpen
  \bibfield  {author} {\bibinfo {author} {\bibfnamefont {S.}~\bibnamefont
  {Acharya}} \emph {et~al.} (\bibinfo {collaboration} {ALICE}),\ }\href
  {\doibase 10.1007/JHEP09(2017)032} {\bibfield  {journal} {\bibinfo  {journal}
  {JHEP}\ }\textbf {\bibinfo {volume} {09}},\ \bibinfo {pages} {032} (\bibinfo
  {year} {2017})},\ \Eprint {http://arxiv.org/abs/1707.05690} {arXiv:1707.05690
  [nucl-ex]} \BibitemShut {NoStop}%
%%CITATION = ARXIV:1707.05690;%%
\bibitem [{\citenamefont {Drescher}\ \emph {et~al.}(2001)\citenamefont
  {Drescher}, \citenamefont {Hladik}, \citenamefont {Ostapchenko},
  \citenamefont {Pierog},\ and\ \citenamefont {Werner}}]{NeXus}%
  \BibitemOpen
  \bibfield  {author} {\bibinfo {author} {\bibfnamefont {H.~J.}\ \bibnamefont
  {Drescher}}, \bibinfo {author} {\bibfnamefont {M.}~\bibnamefont {Hladik}},
  \bibinfo {author} {\bibfnamefont {S.}~\bibnamefont {Ostapchenko}}, \bibinfo
  {author} {\bibfnamefont {T.}~\bibnamefont {Pierog}}, \ and\ \bibinfo {author}
  {\bibfnamefont {K.}~\bibnamefont {Werner}},\ }\href@noop {} {\bibfield
  {journal} {\bibinfo  {journal} {Phys. Rept.}\ }\textbf {\bibinfo {volume}
  {350}},\ \bibinfo {pages} {93} (\bibinfo {year} {2001})},\ \Eprint
  {http://arxiv.org/abs/arXiv:hep-ph/0007198} {arXiv:hep-ph/0007198}
  \BibitemShut {NoStop}%
\bibitem [{\citenamefont {Moreland}\ \emph {et~al.}(2015)\citenamefont
  {Moreland}, \citenamefont {Bernhard},\ and\ \citenamefont
  {Bass}}]{Moreland:2014oya}%
  \BibitemOpen
  \bibfield  {author} {\bibinfo {author} {\bibfnamefont {J.~S.}\ \bibnamefont
  {Moreland}}, \bibinfo {author} {\bibfnamefont {J.~E.}\ \bibnamefont
  {Bernhard}}, \ and\ \bibinfo {author} {\bibfnamefont {S.~A.}\ \bibnamefont
  {Bass}},\ }\href {\doibase 10.1103/PhysRevC.92.011901} {\bibfield  {journal}
  {\bibinfo  {journal} {Phys. Rev.}\ }\textbf {\bibinfo {volume} {C92}},\
  \bibinfo {pages} {011901} (\bibinfo {year} {2015})},\ \Eprint
  {http://arxiv.org/abs/1412.4708} {arXiv:1412.4708 [nucl-th]} \BibitemShut
  {NoStop}%
%%CITATION = ARXIV:1412.4708;%%
\bibitem [{\citenamefont {Miller}\ \emph {et~al.}(2007)\citenamefont {Miller},
  \citenamefont {Reygers}, \citenamefont {Sanders},\ and\ \citenamefont
  {Steinberg}}]{glauber1}%
  \BibitemOpen
  \bibfield  {author} {\bibinfo {author} {\bibfnamefont {M.}~\bibnamefont
  {Miller}}, \bibinfo {author} {\bibfnamefont {K.}~\bibnamefont {Reygers}},
  \bibinfo {author} {\bibfnamefont {S.~J.}\ \bibnamefont {Sanders}}, \ and\
  \bibinfo {author} {\bibfnamefont {P.}~\bibnamefont {Steinberg}},\ }\href@noop
  {} {\bibfield  {journal} {\bibinfo  {journal} {Ann.Rev.Nucl.Part.Sci.}\
  }\textbf {\bibinfo {volume} {57}},\ \bibinfo {pages} {205} (\bibinfo {year}
  {2007})},\ \Eprint {http://arxiv.org/abs/arXiv:nucl-ex/0701025}
  {arXiv:nucl-ex/0701025} \BibitemShut {NoStop}%
\bibitem [{\citenamefont {B.Alver}\ \emph {et~al.}()\citenamefont {B.Alver},
  \citenamefont {M.Baker}, \citenamefont {C.Loizides},\ and\ \citenamefont
  {P.Steinberg}}]{glauber2}%
  \BibitemOpen
  \bibfield  {author} {\bibinfo {author} {\bibnamefont {B.Alver}}, \bibinfo
  {author} {\bibnamefont {M.Baker}}, \bibinfo {author} {\bibnamefont
  {C.Loizides}}, \ and\ \bibinfo {author} {\bibnamefont {P.Steinberg}},\
  }\href@noop {} {}\Eprint {http://arxiv.org/abs/arXiv:0805.4411}
  {arXiv:0805.4411} \BibitemShut {NoStop}%
\bibitem [{\citenamefont {Loizides}\ \emph {et~al.}()\citenamefont {Loizides},
  \citenamefont {Nagle},\ and\ \citenamefont {Steinberg}}]{glauber3}%
  \BibitemOpen
  \bibfield  {author} {\bibinfo {author} {\bibfnamefont {C.}~\bibnamefont
  {Loizides}}, \bibinfo {author} {\bibfnamefont {J.}~\bibnamefont {Nagle}}, \
  and\ \bibinfo {author} {\bibfnamefont {P.}~\bibnamefont {Steinberg}},\
  }\href@noop {} {}\Eprint {http://arxiv.org/abs/arXiv:1408.2549}
  {arXiv:1408.2549} \BibitemShut {NoStop}%
\bibitem [{\citenamefont {H.-J.Drescher}\ and\ \citenamefont
  {Nara}(2007)}]{MCKLN}%
  \BibitemOpen
  \bibfield  {author} {\bibinfo {author} {\bibnamefont {H.-J.Drescher}}\ and\
  \bibinfo {author} {\bibfnamefont {Y.}~\bibnamefont {Nara}},\ }\href@noop {}
  {\bibfield  {journal} {\bibinfo  {journal} {Phys. Rev. C}\ }\textbf {\bibinfo
  {volume} {75}},\ \bibinfo {pages} {034905} (\bibinfo {year} {2007})},\
  \Eprint {http://arxiv.org/abs/arXiv:nucl-th/0611017} {arXiv:nucl-th/0611017}
  \BibitemShut {NoStop}%
\bibitem [{\citenamefont {Renk}\ and\ \citenamefont {Niemi}(2014)}]{Renk14}%
  \BibitemOpen
  \bibfield  {author} {\bibinfo {author} {\bibfnamefont {T.}~\bibnamefont
  {Renk}}\ and\ \bibinfo {author} {\bibfnamefont {H.}~\bibnamefont {Niemi}},\
  }\href {\doibase 10.1103/PhysRevC.89.064907} {\bibfield  {journal} {\bibinfo
  {journal} {Phys. Rev.}\ }\textbf {\bibinfo {volume} {C89}},\ \bibinfo {pages}
  {064907} (\bibinfo {year} {2014})},\ \Eprint {http://arxiv.org/abs/1401.2069}
  {arXiv:1401.2069 [nucl-th]} \BibitemShut {NoStop}%
%%CITATION = ARXIV:1401.2069;%%
\bibitem [{\citenamefont {S.Ghosh}\ \emph {et~al.}(2016)\citenamefont
  {S.Ghosh}, \citenamefont {Singh}, \citenamefont {Chatterjee}, \citenamefont
  {Alam},\ and\ \citenamefont {Sarkar}}]{Ghosh16}%
  \BibitemOpen
  \bibfield  {author} {\bibinfo {author} {\bibnamefont {S.Ghosh}}, \bibinfo
  {author} {\bibfnamefont {S.~K.}\ \bibnamefont {Singh}}, \bibinfo {author}
  {\bibfnamefont {S.}~\bibnamefont {Chatterjee}}, \bibinfo {author}
  {\bibfnamefont {J.}~\bibnamefont {Alam}}, \ and\ \bibinfo {author}
  {\bibfnamefont {S.}~\bibnamefont {Sarkar}},\ }\href@noop {} {\bibfield
  {journal} {\bibinfo  {journal} {Phys. Rev. C}\ }\textbf {\bibinfo {volume}
  {93}},\ \bibinfo {pages} {054904} (\bibinfo {year} {2016})},\ \Eprint
  {http://arxiv.org/abs/1601.03971} {arXiv:1601.03971} \BibitemShut {NoStop}%
\bibitem [{\citenamefont {Schenke}\ \emph {et~al.}(2012)\citenamefont
  {Schenke}, \citenamefont {Tribedy},\ and\ \citenamefont
  {Venugopalan}}]{Schenke12}%
  \BibitemOpen
  \bibfield  {author} {\bibinfo {author} {\bibfnamefont {B.}~\bibnamefont
  {Schenke}}, \bibinfo {author} {\bibfnamefont {P.}~\bibnamefont {Tribedy}}, \
  and\ \bibinfo {author} {\bibfnamefont {R.}~\bibnamefont {Venugopalan}},\
  }\href@noop {} {\bibfield  {journal} {\bibinfo  {journal} {Phys. Rev. Lett.}\
  }\textbf {\bibinfo {volume} {108}},\ \bibinfo {pages} {252301} (\bibinfo
  {year} {2012})},\ \Eprint {http://arxiv.org/abs/arXiv:1202.6646}
  {arXiv:1202.6646} \BibitemShut {NoStop}%
\bibitem [{\citenamefont {McDonald}\ \emph {et~al.}(2017)\citenamefont
  {McDonald}, \citenamefont {Shen}, \citenamefont {Fillion-Gourdeau},
  \citenamefont {Jeon},\ and\ \citenamefont {Gale}}]{McDonald:2016vlt}%
  \BibitemOpen
  \bibfield  {author} {\bibinfo {author} {\bibfnamefont {S.}~\bibnamefont
  {McDonald}}, \bibinfo {author} {\bibfnamefont {C.}~\bibnamefont {Shen}},
  \bibinfo {author} {\bibfnamefont {F.}~\bibnamefont {Fillion-Gourdeau}},
  \bibinfo {author} {\bibfnamefont {S.}~\bibnamefont {Jeon}}, \ and\ \bibinfo
  {author} {\bibfnamefont {C.}~\bibnamefont {Gale}},\ }\href {\doibase
  10.1103/PhysRevC.95.064913} {\bibfield  {journal} {\bibinfo  {journal} {Phys.
  Rev.}\ }\textbf {\bibinfo {volume} {C95}},\ \bibinfo {pages} {064913}
  (\bibinfo {year} {2017})},\ \Eprint {http://arxiv.org/abs/1609.02958}
  {arXiv:1609.02958 [hep-ph]} \BibitemShut {NoStop}%
%%CITATION = ARXIV:1609.02958;%%
\bibitem [{\citenamefont {Gale}\ \emph
  {et~al.}(2013{\natexlab{b}})\citenamefont {Gale}, \citenamefont {Jeon},
  \citenamefont {Schenke}, \citenamefont {Tribedy},\ and\ \citenamefont
  {Venugopalan}}]{Schenke13}%
  \BibitemOpen
  \bibfield  {author} {\bibinfo {author} {\bibfnamefont {C.}~\bibnamefont
  {Gale}}, \bibinfo {author} {\bibfnamefont {S.}~\bibnamefont {Jeon}}, \bibinfo
  {author} {\bibfnamefont {B.}~\bibnamefont {Schenke}}, \bibinfo {author}
  {\bibfnamefont {P.}~\bibnamefont {Tribedy}}, \ and\ \bibinfo {author}
  {\bibfnamefont {R.}~\bibnamefont {Venugopalan}},\ }\href@noop {} {\bibfield
  {journal} {\bibinfo  {journal} {Phys. Rev. Lett.}\ }\textbf {\bibinfo
  {volume} {110}},\ \bibinfo {pages} {012302} (\bibinfo {year}
  {2013}{\natexlab{b}})},\ \Eprint {http://arxiv.org/abs/arXiv:1209.6330}
  {arXiv:1209.6330} \BibitemShut {NoStop}%
\bibitem [{\citenamefont {Eskola}\ \emph {et~al.}(2000)\citenamefont {Eskola},
  \citenamefont {Kajantie}, \citenamefont {Ruuskanen},\ and\ \citenamefont
  {Tuominen}}]{EKRT}%
  \BibitemOpen
  \bibfield  {author} {\bibinfo {author} {\bibfnamefont {K.~J.}\ \bibnamefont
  {Eskola}}, \bibinfo {author} {\bibfnamefont {K.}~\bibnamefont {Kajantie}},
  \bibinfo {author} {\bibfnamefont {P.~V.}\ \bibnamefont {Ruuskanen}}, \ and\
  \bibinfo {author} {\bibfnamefont {K.}~\bibnamefont {Tuominen}},\ }\href@noop
  {} {\bibfield  {journal} {\bibinfo  {journal} {Nucl. Phys.}\ }\textbf
  {\bibinfo {volume} {B570}},\ \bibinfo {pages} {379} (\bibinfo {year}
  {2000})},\ \Eprint {http://arxiv.org/abs/arXiv:hep-ph/9909456}
  {arXiv:hep-ph/9909456} \BibitemShut {NoStop}%
\bibitem [{\citenamefont {Niemi}\ \emph {et~al.}(2016)\citenamefont {Niemi},
  \citenamefont {Eskola},\ and\ \citenamefont {Paatelainen}}]{Niemi15}%
  \BibitemOpen
  \bibfield  {author} {\bibinfo {author} {\bibfnamefont {H.}~\bibnamefont
  {Niemi}}, \bibinfo {author} {\bibfnamefont {K.~J.}\ \bibnamefont {Eskola}}, \
  and\ \bibinfo {author} {\bibfnamefont {R.}~\bibnamefont {Paatelainen}},\
  }\href {\doibase 10.1103/PhysRevC.93.024907} {\bibfield  {journal} {\bibinfo
  {journal} {Phys. Rev.}\ }\textbf {\bibinfo {volume} {C93}},\ \bibinfo {pages}
  {024907} (\bibinfo {year} {2016})},\ \Eprint
  {http://arxiv.org/abs/1505.02677} {arXiv:1505.02677 [hep-ph]} \BibitemShut
  {NoStop}%
%%CITATION = ARXIV:1505.02677;%%
\bibitem [{\citenamefont {Zhang}\ \emph {et~al.}(2000)\citenamefont {Zhang},
  \citenamefont {Ko}, \citenamefont {Li},\ and\ \citenamefont
  {Lin}}]{Zhang:1999bd}%
  \BibitemOpen
  \bibfield  {author} {\bibinfo {author} {\bibfnamefont {B.}~\bibnamefont
  {Zhang}}, \bibinfo {author} {\bibfnamefont {C.~M.}\ \bibnamefont {Ko}},
  \bibinfo {author} {\bibfnamefont {B.-A.}\ \bibnamefont {Li}}, \ and\ \bibinfo
  {author} {\bibfnamefont {Z.-w.}\ \bibnamefont {Lin}},\ }\href {\doibase
  10.1103/PhysRevC.61.067901} {\bibfield  {journal} {\bibinfo  {journal} {Phys.
  Rev.}\ }\textbf {\bibinfo {volume} {C61}},\ \bibinfo {pages} {067901}
  (\bibinfo {year} {2000})},\ \Eprint {http://arxiv.org/abs/nucl-th/9907017}
  {arXiv:nucl-th/9907017 [nucl-th]} \BibitemShut {NoStop}%
%%CITATION = NUCL-TH/9907017;%%
\bibitem [{\citenamefont {Zhao}\ \emph {et~al.}(2017)\citenamefont {Zhao},
  \citenamefont {Xu},\ and\ \citenamefont {Song}}]{Zhao:2017yhj}%
  \BibitemOpen
  \bibfield  {author} {\bibinfo {author} {\bibfnamefont {W.}~\bibnamefont
  {Zhao}}, \bibinfo {author} {\bibfnamefont {H.-j.}\ \bibnamefont {Xu}}, \ and\
  \bibinfo {author} {\bibfnamefont {H.}~\bibnamefont {Song}},\ }\href {\doibase
  10.1140/epjc/s10052-017-5186-x} {\bibfield  {journal} {\bibinfo  {journal}
  {Eur. Phys. J.}\ }\textbf {\bibinfo {volume} {C77}},\ \bibinfo {pages} {645}
  (\bibinfo {year} {2017})},\ \Eprint {http://arxiv.org/abs/1703.10792}
  {arXiv:1703.10792 [nucl-th]} \BibitemShut {NoStop}%
%%CITATION = ARXIV:1703.10792;%%
\bibitem [{\citenamefont {Niemi}\ \emph {et~al.}(2013)\citenamefont {Niemi},
  \citenamefont {Denicol}, \citenamefont {Holopainen},\ and\ \citenamefont
  {Huovinen}}]{Niemi12}%
  \BibitemOpen
  \bibfield  {author} {\bibinfo {author} {\bibfnamefont {H.}~\bibnamefont
  {Niemi}}, \bibinfo {author} {\bibfnamefont {G.~S.}\ \bibnamefont {Denicol}},
  \bibinfo {author} {\bibfnamefont {H.}~\bibnamefont {Holopainen}}, \ and\
  \bibinfo {author} {\bibfnamefont {P.}~\bibnamefont {Huovinen}},\ }\href@noop
  {} {\bibfield  {journal} {\bibinfo  {journal} {Phys. Rev. C}\ }\textbf
  {\bibinfo {volume} {87}},\ \bibinfo {pages} {054901} (\bibinfo {year}
  {2013})},\ \Eprint {http://arxiv.org/abs/arXiv:1212.1008} {arXiv:1212.1008}
  \BibitemShut {NoStop}%
\bibitem [{\citenamefont {Gardim}\ \emph {et~al.}(2018)\citenamefont {Gardim},
  \citenamefont {Grassi}, \citenamefont {Ishida}, \citenamefont {Luzum},
  \citenamefont {Magalh\~aes},\ and\ \citenamefont
  {Noronha-Hostler}}]{Gardim:2017ruc}%
  \BibitemOpen
  \bibfield  {author} {\bibinfo {author} {\bibfnamefont {F.~G.}\ \bibnamefont
  {Gardim}}, \bibinfo {author} {\bibfnamefont {F.}~\bibnamefont {Grassi}},
  \bibinfo {author} {\bibfnamefont {P.}~\bibnamefont {Ishida}}, \bibinfo
  {author} {\bibfnamefont {M.}~\bibnamefont {Luzum}}, \bibinfo {author}
  {\bibfnamefont {P.~S.}\ \bibnamefont {Magalh\~aes}}, \ and\ \bibinfo {author}
  {\bibfnamefont {J.}~\bibnamefont {Noronha-Hostler}},\ }\href {\doibase
  10.1103/PhysRevC.97.064919} {\bibfield  {journal} {\bibinfo  {journal} {Phys.
  Rev. C}\ }\textbf {\bibinfo {volume} {97}},\ \bibinfo {pages} {064919}
  (\bibinfo {year} {2018})},\ \Eprint {http://arxiv.org/abs/1712.03912}
  {arXiv:1712.03912 [nucl-th]} \BibitemShut {NoStop}%
\bibitem [{\citenamefont {Gardim}\ \emph
  {et~al.}(2012{\natexlab{a}})\citenamefont {Gardim}, \citenamefont {Grassi},
  \citenamefont {Luzum},\ and\ \citenamefont {Ollitrault}}]{sph1}%
  \BibitemOpen
  \bibfield  {author} {\bibinfo {author} {\bibfnamefont {F.~G.}\ \bibnamefont
  {Gardim}}, \bibinfo {author} {\bibfnamefont {F.}~\bibnamefont {Grassi}},
  \bibinfo {author} {\bibfnamefont {M.}~\bibnamefont {Luzum}}, \ and\ \bibinfo
  {author} {\bibfnamefont {J.-Y.}\ \bibnamefont {Ollitrault}},\ }\href@noop {}
  {\bibfield  {journal} {\bibinfo  {journal} {Phys. Rev. C}\ }\textbf {\bibinfo
  {volume} {85}},\ \bibinfo {pages} {024908} (\bibinfo {year}
  {2012}{\natexlab{a}})},\ \Eprint {http://arxiv.org/abs/arXiv:1111.6538}
  {arXiv:1111.6538} \BibitemShut {NoStop}%
\bibitem [{\citenamefont {Gardim}\ \emph {et~al.}(2015)\citenamefont {Gardim},
  \citenamefont {Noronha-Hostler}, \citenamefont {Luzum},\ and\ \citenamefont
  {Grassi}}]{sph2}%
  \BibitemOpen
  \bibfield  {author} {\bibinfo {author} {\bibfnamefont {F.~G.}\ \bibnamefont
  {Gardim}}, \bibinfo {author} {\bibfnamefont {J.}~\bibnamefont
  {Noronha-Hostler}}, \bibinfo {author} {\bibfnamefont {M.}~\bibnamefont
  {Luzum}}, \ and\ \bibinfo {author} {\bibfnamefont {F.}~\bibnamefont
  {Grassi}},\ }\href@noop {} {\bibfield  {journal} {\bibinfo  {journal} {Phys.
  Rev. C}\ }\textbf {\bibinfo {volume} {91}},\ \bibinfo {pages} {034902}
  (\bibinfo {year} {2015})},\ \Eprint {http://arxiv.org/abs/arXiv:1411.2574}
  {arXiv:1411.2574} \BibitemShut {NoStop}%
\bibitem [{\citenamefont {Fu}(2015)}]{Fu15}%
  \BibitemOpen
  \bibfield  {author} {\bibinfo {author} {\bibfnamefont {J.}~\bibnamefont
  {Fu}},\ }\href@noop {} {\bibfield  {journal} {\bibinfo  {journal} {Phys. Rev.
  C}\ }\textbf {\bibinfo {volume} {92}},\ \bibinfo {pages} {024904} (\bibinfo
  {year} {2015})}\BibitemShut {NoStop}%
\bibitem [{\citenamefont {Noronha-Hostler}\ \emph
  {et~al.}(2016{\natexlab{a}})\citenamefont {Noronha-Hostler}, \citenamefont
  {Betz}, \citenamefont {Noronha},\ and\ \citenamefont
  {Gyulassy}}]{Noronha-Hostler:2016eow}%
  \BibitemOpen
  \bibfield  {author} {\bibinfo {author} {\bibfnamefont {J.}~\bibnamefont
  {Noronha-Hostler}}, \bibinfo {author} {\bibfnamefont {B.}~\bibnamefont
  {Betz}}, \bibinfo {author} {\bibfnamefont {J.}~\bibnamefont {Noronha}}, \
  and\ \bibinfo {author} {\bibfnamefont {M.}~\bibnamefont {Gyulassy}},\ }\href
  {\doibase 10.1103/PhysRevLett.116.252301} {\bibfield  {journal} {\bibinfo
  {journal} {Phys. Rev. Lett.}\ }\textbf {\bibinfo {volume} {116}},\ \bibinfo
  {pages} {252301} (\bibinfo {year} {2016}{\natexlab{a}})},\ \Eprint
  {http://arxiv.org/abs/1602.03788} {arXiv:1602.03788 [nucl-th]} \BibitemShut
  {NoStop}%
%%CITATION = ARXIV:1602.03788;%%
\bibitem [{\citenamefont {Betz}\ \emph {et~al.}(2017)\citenamefont {Betz},
  \citenamefont {Gyulassy}, \citenamefont {Luzum}, \citenamefont {Noronha},
  \citenamefont {Noronha-Hostler}, \citenamefont {Portillo},\ and\
  \citenamefont {Ratti}}]{Betz:2016ayq}%
  \BibitemOpen
  \bibfield  {author} {\bibinfo {author} {\bibfnamefont {B.}~\bibnamefont
  {Betz}}, \bibinfo {author} {\bibfnamefont {M.}~\bibnamefont {Gyulassy}},
  \bibinfo {author} {\bibfnamefont {M.}~\bibnamefont {Luzum}}, \bibinfo
  {author} {\bibfnamefont {J.}~\bibnamefont {Noronha}}, \bibinfo {author}
  {\bibfnamefont {J.}~\bibnamefont {Noronha-Hostler}}, \bibinfo {author}
  {\bibfnamefont {I.}~\bibnamefont {Portillo}}, \ and\ \bibinfo {author}
  {\bibfnamefont {C.}~\bibnamefont {Ratti}},\ }\href {\doibase
  10.1103/PhysRevC.95.044901} {\bibfield  {journal} {\bibinfo  {journal} {Phys.
  Rev.}\ }\textbf {\bibinfo {volume} {C95}},\ \bibinfo {pages} {044901}
  (\bibinfo {year} {2017})},\ \Eprint {http://arxiv.org/abs/1609.05171}
  {arXiv:1609.05171 [nucl-th]} \BibitemShut {NoStop}%
%%CITATION = ARXIV:1609.05171;%%
\bibitem [{\citenamefont {Prado}\ \emph {et~al.}(2016)\citenamefont {Prado},
  \citenamefont {Noronha-Hostler}, \citenamefont {Suaide}, \citenamefont
  {Noronha}, \citenamefont {Munhoz},\ and\ \citenamefont
  {Cosentino}}]{Prado:2016szr}%
  \BibitemOpen
  \bibfield  {author} {\bibinfo {author} {\bibfnamefont {C.~A.~G.}\
  \bibnamefont {Prado}}, \bibinfo {author} {\bibfnamefont {J.}~\bibnamefont
  {Noronha-Hostler}}, \bibinfo {author} {\bibfnamefont {A.~A.~P.}\ \bibnamefont
  {Suaide}}, \bibinfo {author} {\bibfnamefont {J.}~\bibnamefont {Noronha}},
  \bibinfo {author} {\bibfnamefont {M.~G.}\ \bibnamefont {Munhoz}}, \ and\
  \bibinfo {author} {\bibfnamefont {M.~R.}\ \bibnamefont {Cosentino}},\
  }\href@noop {} {\  (\bibinfo {year} {2016})},\ \Eprint
  {http://arxiv.org/abs/1611.02965} {arXiv:1611.02965 [nucl-th]} \BibitemShut
  {NoStop}%
%%CITATION = ARXIV:1611.02965;%%
\bibitem [{\citenamefont {Katz}\ \emph {et~al.}(2020)\citenamefont {Katz},
  \citenamefont {Prado}, \citenamefont {Noronha-Hostler}, \citenamefont
  {Noronha},\ and\ \citenamefont {Suaide}}]{Katz:2019fkc}%
  \BibitemOpen
  \bibfield  {author} {\bibinfo {author} {\bibfnamefont {R.}~\bibnamefont
  {Katz}}, \bibinfo {author} {\bibfnamefont {C.~A.~G.}\ \bibnamefont {Prado}},
  \bibinfo {author} {\bibfnamefont {J.}~\bibnamefont {Noronha-Hostler}},
  \bibinfo {author} {\bibfnamefont {J.}~\bibnamefont {Noronha}}, \ and\
  \bibinfo {author} {\bibfnamefont {A.~A.~P.}\ \bibnamefont {Suaide}},\ }\href
  {\doibase 10.1103/PhysRevC.102.024906} {\bibfield  {journal} {\bibinfo
  {journal} {Phys. Rev. C}\ }\textbf {\bibinfo {volume} {102}},\ \bibinfo
  {pages} {024906} (\bibinfo {year} {2020})},\ \Eprint
  {http://arxiv.org/abs/1906.10768} {arXiv:1906.10768 [nucl-th]} \BibitemShut
  {NoStop}%
\bibitem [{\citenamefont {Noronha-Hostler}\ \emph
  {et~al.}(2016{\natexlab{b}})\citenamefont {Noronha-Hostler}, \citenamefont
  {Yan}, \citenamefont {Gardim},\ and\ \citenamefont {Ollitraul}}]{Jaki15}%
  \BibitemOpen
  \bibfield  {author} {\bibinfo {author} {\bibfnamefont {J.}~\bibnamefont
  {Noronha-Hostler}}, \bibinfo {author} {\bibfnamefont {L.}~\bibnamefont
  {Yan}}, \bibinfo {author} {\bibfnamefont {F.~G.}\ \bibnamefont {Gardim}}, \
  and\ \bibinfo {author} {\bibfnamefont {J.-Y.}\ \bibnamefont {Ollitraul}},\
  }\href@noop {} {\bibfield  {journal} {\bibinfo  {journal} {Phys. Rev. C}\
  }\textbf {\bibinfo {volume} {93}},\ \bibinfo {pages} {014909} (\bibinfo
  {year} {2016}{\natexlab{b}})},\ \Eprint
  {http://arxiv.org/abs/arXiv:1511.03896} {arXiv:1511.03896} \BibitemShut
  {NoStop}%
\bibitem [{\citenamefont {Sievert}\ and\ \citenamefont
  {Noronha-Hostler}(2019)}]{Sievert:2019zjr}%
  \BibitemOpen
  \bibfield  {author} {\bibinfo {author} {\bibfnamefont {M.~D.}\ \bibnamefont
  {Sievert}}\ and\ \bibinfo {author} {\bibfnamefont {J.}~\bibnamefont
  {Noronha-Hostler}},\ }\href {\doibase 10.1103/PhysRevC.100.024904} {\bibfield
   {journal} {\bibinfo  {journal} {Phys. Rev. C}\ }\textbf {\bibinfo {volume}
  {100}},\ \bibinfo {pages} {024904} (\bibinfo {year} {2019})},\ \Eprint
  {http://arxiv.org/abs/1901.01319} {arXiv:1901.01319 [nucl-th]} \BibitemShut
  {NoStop}%
\bibitem [{\citenamefont {Rao}\ \emph {et~al.}(2021)\citenamefont {Rao},
  \citenamefont {Sievert},\ and\ \citenamefont
  {Noronha-Hostler}}]{Rao:2019vgy}%
  \BibitemOpen
  \bibfield  {author} {\bibinfo {author} {\bibfnamefont {S.}~\bibnamefont
  {Rao}}, \bibinfo {author} {\bibfnamefont {M.}~\bibnamefont {Sievert}}, \ and\
  \bibinfo {author} {\bibfnamefont {J.}~\bibnamefont {Noronha-Hostler}},\
  }\href {\doibase 10.1103/PhysRevC.103.034910} {\bibfield  {journal} {\bibinfo
   {journal} {Phys. Rev. C}\ }\textbf {\bibinfo {volume} {103}},\ \bibinfo
  {pages} {034910} (\bibinfo {year} {2021})},\ \Eprint
  {http://arxiv.org/abs/1910.03677} {arXiv:1910.03677 [nucl-th]} \BibitemShut
  {NoStop}%
\bibitem [{\citenamefont {Hippert}\ \emph {et~al.}(2020)\citenamefont
  {Hippert}, \citenamefont {Barbon}, \citenamefont {Dobrigkeit~Chinellato},
  \citenamefont {Luzum}, \citenamefont {Noronha}, \citenamefont {Nunes~da
  Silva}, \citenamefont {Serenone},\ and\ \citenamefont
  {Takahashi}}]{Hippert:2020kde}%
  \BibitemOpen
  \bibfield  {author} {\bibinfo {author} {\bibfnamefont {M.}~\bibnamefont
  {Hippert}}, \bibinfo {author} {\bibfnamefont {J.~a. G.~P.}\ \bibnamefont
  {Barbon}}, \bibinfo {author} {\bibfnamefont {D.}~\bibnamefont
  {Dobrigkeit~Chinellato}}, \bibinfo {author} {\bibfnamefont {M.}~\bibnamefont
  {Luzum}}, \bibinfo {author} {\bibfnamefont {J.}~\bibnamefont {Noronha}},
  \bibinfo {author} {\bibfnamefont {T.}~\bibnamefont {Nunes~da Silva}},
  \bibinfo {author} {\bibfnamefont {W.~M.}\ \bibnamefont {Serenone}}, \ and\
  \bibinfo {author} {\bibfnamefont {J.}~\bibnamefont {Takahashi}},\ }\href
  {\doibase 10.1103/PhysRevC.102.064909} {\bibfield  {journal} {\bibinfo
  {journal} {Phys. Rev. C}\ }\textbf {\bibinfo {volume} {102}},\ \bibinfo
  {pages} {064909} (\bibinfo {year} {2020})},\ \Eprint
  {http://arxiv.org/abs/2006.13358} {arXiv:2006.13358 [nucl-th]} \BibitemShut
  {NoStop}%
\bibitem [{\citenamefont {Voloshin}\ \emph {et~al.}(2008)\citenamefont
  {Voloshin}, \citenamefont {Poskanzer}, \citenamefont {Tang},\ and\
  \citenamefont {Wang}}]{Voloshin:2007pc}%
  \BibitemOpen
  \bibfield  {author} {\bibinfo {author} {\bibfnamefont {S.~A.}\ \bibnamefont
  {Voloshin}}, \bibinfo {author} {\bibfnamefont {A.~M.}\ \bibnamefont
  {Poskanzer}}, \bibinfo {author} {\bibfnamefont {A.}~\bibnamefont {Tang}}, \
  and\ \bibinfo {author} {\bibfnamefont {G.}~\bibnamefont {Wang}},\ }\href
  {\doibase 10.1016/j.physletb.2007.11.043} {\bibfield  {journal} {\bibinfo
  {journal} {Phys. Lett.}\ }\textbf {\bibinfo {volume} {B659}},\ \bibinfo
  {pages} {537} (\bibinfo {year} {2008})},\ \Eprint
  {http://arxiv.org/abs/0708.0800} {arXiv:0708.0800 [nucl-th]} \BibitemShut
  {NoStop}%
%%CITATION = ARXIV:0708.0800;%%
\bibitem [{\citenamefont {Yan}\ and\ \citenamefont {Ollitrault}(2014)}]{JYPRL}%
  \BibitemOpen
  \bibfield  {author} {\bibinfo {author} {\bibfnamefont {L.}~\bibnamefont
  {Yan}}\ and\ \bibinfo {author} {\bibfnamefont {J.-Y.}\ \bibnamefont
  {Ollitrault}},\ }\href@noop {} {\bibfield  {journal} {\bibinfo  {journal}
  {Phys. Rev. Lett.}\ }\textbf {\bibinfo {volume} {112}},\ \bibinfo {pages}
  {082301} (\bibinfo {year} {2014})},\ \Eprint
  {http://arxiv.org/abs/arXiv:1312.6555} {arXiv:1312.6555} \BibitemShut
  {NoStop}%
\bibitem [{\citenamefont {Yan}\ \emph {et~al.}(2014)\citenamefont {Yan},
  \citenamefont {Ollitrault},\ and\ \citenamefont {Poskanzer}}]{JYPRC}%
  \BibitemOpen
  \bibfield  {author} {\bibinfo {author} {\bibfnamefont {L.}~\bibnamefont
  {Yan}}, \bibinfo {author} {\bibfnamefont {J.-Y.}\ \bibnamefont {Ollitrault}},
  \ and\ \bibinfo {author} {\bibfnamefont {A.}~\bibnamefont {Poskanzer}},\
  }\href@noop {} {\bibfield  {journal} {\bibinfo  {journal} {Phys. Rev. C}\
  }\textbf {\bibinfo {volume} {90}},\ \bibinfo {pages} {024903} (\bibinfo
  {year} {2014})},\ \Eprint {http://arxiv.org/abs/arXiv:1405.6595}
  {arXiv:1405.6595} \BibitemShut {NoStop}%
\bibitem [{\citenamefont {Yan}\ \emph {et~al.}(2015)\citenamefont {Yan},
  \citenamefont {Ollitrault},\ and\ \citenamefont {Poskanzer}}]{JYPLB}%
  \BibitemOpen
  \bibfield  {author} {\bibinfo {author} {\bibfnamefont {L.}~\bibnamefont
  {Yan}}, \bibinfo {author} {\bibfnamefont {J.-Y.}\ \bibnamefont {Ollitrault}},
  \ and\ \bibinfo {author} {\bibfnamefont {A.}~\bibnamefont {Poskanzer}},\
  }\href@noop {} {\bibfield  {journal} {\bibinfo  {journal} {Phys. Lett. B}\
  }\textbf {\bibinfo {volume} {742}},\ \bibinfo {pages} {290} (\bibinfo {year}
  {2015})},\ \Eprint {http://arxiv.org/abs/arXiv:1408.0921} {arXiv:1408.0921}
  \BibitemShut {NoStop}%
\bibitem [{\citenamefont {Gardim}\ and\ \citenamefont
  {Ollitrault}(2021)}]{Gardim:2020mmy}%
  \BibitemOpen
  \bibfield  {author} {\bibinfo {author} {\bibfnamefont {F.~G.}\ \bibnamefont
  {Gardim}}\ and\ \bibinfo {author} {\bibfnamefont {J.-Y.}\ \bibnamefont
  {Ollitrault}},\ }\href {\doibase 10.1103/PhysRevC.103.044907} {\bibfield
  {journal} {\bibinfo  {journal} {Phys. Rev. C}\ }\textbf {\bibinfo {volume}
  {103}},\ \bibinfo {pages} {044907} (\bibinfo {year} {2021})},\ \Eprint
  {http://arxiv.org/abs/2010.11919} {arXiv:2010.11919 [nucl-th]} \BibitemShut
  {NoStop}%
\bibitem [{\citenamefont {F.Gardim}\ \emph {et~al.}(2012)\citenamefont
  {F.Gardim}, \citenamefont {F.Grassi}, \citenamefont {M.Luzum},\ and\
  \citenamefont {J.-Y.Ollitrault}}]{gardim13}%
  \BibitemOpen
  \bibfield  {author} {\bibinfo {author} {\bibnamefont {F.Gardim}}, \bibinfo
  {author} {\bibnamefont {F.Grassi}}, \bibinfo {author} {\bibnamefont
  {M.Luzum}}, \ and\ \bibinfo {author} {\bibnamefont {J.-Y.Ollitrault}},\
  }\href@noop {} {\bibfield  {journal} {\bibinfo  {journal} {Phys. Rev. C}\
  }\textbf {\bibinfo {volume} {87}},\ \bibinfo {pages} {031901(R)} (\bibinfo
  {year} {2012})}\BibitemShut {NoStop}%
\bibitem [{\citenamefont {I.Kozlov}\ \emph {et~al.}()\citenamefont {I.Kozlov},
  \citenamefont {M.Luzum}, \citenamefont {G.Denicol}, \citenamefont {Jeon},\
  and\ \citenamefont {Gale}}]{kozlov14}%
  \BibitemOpen
  \bibfield  {author} {\bibinfo {author} {\bibnamefont {I.Kozlov}}, \bibinfo
  {author} {\bibnamefont {M.Luzum}}, \bibinfo {author} {\bibnamefont
  {G.Denicol}}, \bibinfo {author} {\bibfnamefont {S.}~\bibnamefont {Jeon}}, \
  and\ \bibinfo {author} {\bibfnamefont {C.}~\bibnamefont {Gale}},\ }\href@noop
  {} {\ }\Eprint {http://arxiv.org/abs/arXiv:1405.3976} {arXiv:1405.3976}
  \BibitemShut {NoStop}%
\bibitem [{\citenamefont {U.Heinz}\ \emph {et~al.}(2013)\citenamefont
  {U.Heinz}, \citenamefont {Qiu},\ and\ \citenamefont {Shen}}]{heinz15}%
  \BibitemOpen
  \bibfield  {author} {\bibinfo {author} {\bibnamefont {U.Heinz}}, \bibinfo
  {author} {\bibfnamefont {Z.}~\bibnamefont {Qiu}}, \ and\ \bibinfo {author}
  {\bibfnamefont {C.}~\bibnamefont {Shen}},\ }\href@noop {} {\bibfield
  {journal} {\bibinfo  {journal} {Phys. Rev. C}\ }\textbf {\bibinfo {volume}
  {87}},\ \bibinfo {pages} {034913} (\bibinfo {year} {2013})}\BibitemShut
  {NoStop}%
\bibitem [{\citenamefont {Werner}(1993)}]{Werner:1993uh}%
  \BibitemOpen
  \bibfield  {author} {\bibinfo {author} {\bibfnamefont {K.}~\bibnamefont
  {Werner}},\ }\href {\doibase 10.1016/0370-1573(93)90078-R} {\bibfield
  {journal} {\bibinfo  {journal} {Phys. Rept.}\ }\textbf {\bibinfo {volume}
  {232}},\ \bibinfo {pages} {87} (\bibinfo {year} {1993})}\BibitemShut
  {NoStop}%
\bibitem [{\citenamefont {Drescher}\ \emph {et~al.}(2002)\citenamefont
  {Drescher}, \citenamefont {Ostapchenko}, \citenamefont {Pierog},\ and\
  \citenamefont {Werner}}]{Drescher:2000ec}%
  \BibitemOpen
  \bibfield  {author} {\bibinfo {author} {\bibfnamefont {H.~J.}\ \bibnamefont
  {Drescher}}, \bibinfo {author} {\bibfnamefont {S.}~\bibnamefont
  {Ostapchenko}}, \bibinfo {author} {\bibfnamefont {T.}~\bibnamefont {Pierog}},
  \ and\ \bibinfo {author} {\bibfnamefont {K.}~\bibnamefont {Werner}},\ }\href
  {\doibase 10.1103/PhysRevC.65.054902} {\bibfield  {journal} {\bibinfo
  {journal} {Phys. Rev. C}\ }\textbf {\bibinfo {volume} {65}},\ \bibinfo
  {pages} {054902} (\bibinfo {year} {2002})},\ \Eprint
  {http://arxiv.org/abs/hep-ph/0011219} {arXiv:hep-ph/0011219} \BibitemShut
  {NoStop}%
\bibitem [{\citenamefont {Qian}\ \emph {et~al.}(2007)\citenamefont {Qian},
  \citenamefont {Andrade}, \citenamefont {Grassi}, \citenamefont
  {O.~Socolowski}, \citenamefont {Kodama},\ and\ \citenamefont
  {Hama}}]{NeXspectra}%
  \BibitemOpen
  \bibfield  {author} {\bibinfo {author} {\bibfnamefont {W.~L.}\ \bibnamefont
  {Qian}}, \bibinfo {author} {\bibfnamefont {R.}~\bibnamefont {Andrade}},
  \bibinfo {author} {\bibfnamefont {F.}~\bibnamefont {Grassi}}, \bibinfo
  {author} {\bibfnamefont {J.}~\bibnamefont {O.~Socolowski}}, \bibinfo {author}
  {\bibfnamefont {T.}~\bibnamefont {Kodama}}, \ and\ \bibinfo {author}
  {\bibfnamefont {Y.}~\bibnamefont {Hama}},\ }\href@noop {} {\bibfield
  {journal} {\bibinfo  {journal} {Int. J. Mod. Phys. E}\ }\textbf {\bibinfo
  {volume} {16}},\ \bibinfo {pages} {1877} (\bibinfo {year} {2007})},\ \Eprint
  {http://arxiv.org/abs/nucl-th/arXiv:0703078} {nucl-th/arXiv:0703078}
  \BibitemShut {NoStop}%
\bibitem [{\citenamefont {Andrade}\ \emph {et~al.}(2006)\citenamefont
  {Andrade}, \citenamefont {Grassi}, \citenamefont {Hama}, \citenamefont
  {Kodama},\ and\ \citenamefont {O.~Socolowski}}]{Andrade06}%
  \BibitemOpen
  \bibfield  {author} {\bibinfo {author} {\bibfnamefont {R.~P.~G.}\
  \bibnamefont {Andrade}}, \bibinfo {author} {\bibfnamefont {F.}~\bibnamefont
  {Grassi}}, \bibinfo {author} {\bibfnamefont {Y.}~\bibnamefont {Hama}},
  \bibinfo {author} {\bibfnamefont {T.}~\bibnamefont {Kodama}}, \ and\ \bibinfo
  {author} {\bibfnamefont {J.}~\bibnamefont {O.~Socolowski}},\ }\href@noop {}
  {\bibfield  {journal} {\bibinfo  {journal} {Phys. Rev. Lett.}\ }\textbf
  {\bibinfo {volume} {97}},\ \bibinfo {pages} {202302} (\bibinfo {year}
  {2006})},\ \Eprint {http://arxiv.org/abs/arXiv:nucl-th/0608067}
  {arXiv:nucl-th/0608067} \BibitemShut {NoStop}%
\bibitem [{\citenamefont {Andrade}\ \emph {et~al.}(2008)\citenamefont
  {Andrade}, \citenamefont {Grassi}, \citenamefont {Hama}, \citenamefont
  {Kodama},\ and\ \citenamefont {Qian}}]{Andrade08a}%
  \BibitemOpen
  \bibfield  {author} {\bibinfo {author} {\bibfnamefont {R.~P.~G.}\
  \bibnamefont {Andrade}}, \bibinfo {author} {\bibfnamefont {F.}~\bibnamefont
  {Grassi}}, \bibinfo {author} {\bibfnamefont {Y.}~\bibnamefont {Hama}},
  \bibinfo {author} {\bibfnamefont {T.}~\bibnamefont {Kodama}}, \ and\ \bibinfo
  {author} {\bibfnamefont {W.~L.}\ \bibnamefont {Qian}},\ }\href@noop {}
  {\bibfield  {journal} {\bibinfo  {journal} {Phys. Rev. Lett.}\ }\textbf
  {\bibinfo {volume} {101}},\ \bibinfo {pages} {112301} (\bibinfo {year}
  {2008})},\ \Eprint {http://arxiv.org/abs/arXiv:0805.0018} {arXiv:0805.0018}
  \BibitemShut {NoStop}%
\bibitem [{\citenamefont {Andrade}\ \emph {et~al.}(2009)\citenamefont
  {Andrade}, \citenamefont {dos Reis}, \citenamefont {Grassi}, \citenamefont
  {Hama}, \citenamefont {Qian}, \citenamefont {Kodama},\ and\ \citenamefont
  {Ollitrault}}]{Andrade08b}%
  \BibitemOpen
  \bibfield  {author} {\bibinfo {author} {\bibfnamefont {R.~P.~G.}\
  \bibnamefont {Andrade}}, \bibinfo {author} {\bibfnamefont {A.}~\bibnamefont
  {dos Reis}}, \bibinfo {author} {\bibfnamefont {F.}~\bibnamefont {Grassi}},
  \bibinfo {author} {\bibfnamefont {Y.}~\bibnamefont {Hama}}, \bibinfo {author}
  {\bibfnamefont {W.}~\bibnamefont {Qian}}, \bibinfo {author} {\bibfnamefont
  {T.}~\bibnamefont {Kodama}}, \ and\ \bibinfo {author} {\bibfnamefont {J.-Y.}\
  \bibnamefont {Ollitrault}},\ }\href@noop {} {\bibfield  {journal} {\bibinfo
  {journal} {Acta Phys.Polon.B}\ }\textbf {\bibinfo {volume} {40}},\ \bibinfo
  {pages} {993} (\bibinfo {year} {2009})},\ \Eprint
  {http://arxiv.org/abs/arXiv:0812.4143} {arXiv:0812.4143} \BibitemShut
  {NoStop}%
\bibitem [{\citenamefont {Gardim}\ \emph {et~al.}(2011)\citenamefont {Gardim},
  \citenamefont {Grassi}, \citenamefont {Hama}, \citenamefont {Luzum},\ and\
  \citenamefont {Ollitrault}}]{Gardim11}%
  \BibitemOpen
  \bibfield  {author} {\bibinfo {author} {\bibfnamefont {F.~G.}\ \bibnamefont
  {Gardim}}, \bibinfo {author} {\bibfnamefont {F.}~\bibnamefont {Grassi}},
  \bibinfo {author} {\bibfnamefont {Y.}~\bibnamefont {Hama}}, \bibinfo {author}
  {\bibfnamefont {M.}~\bibnamefont {Luzum}}, \ and\ \bibinfo {author}
  {\bibfnamefont {J.~Y.}\ \bibnamefont {Ollitrault}},\ }\href@noop {}
  {\bibfield  {journal} {\bibinfo  {journal} {Phys. Rev. C}\ }\textbf {\bibinfo
  {volume} {83}},\ \bibinfo {pages} {064901} (\bibinfo {year} {2011})},\
  \Eprint {http://arxiv.org/abs/arXiv:1103.4605} {arXiv:1103.4605} \BibitemShut
  {NoStop}%
\bibitem [{\citenamefont {Gardim}\ \emph
  {et~al.}(2012{\natexlab{b}})\citenamefont {Gardim}, \citenamefont {Grassi},
  \citenamefont {Luzum},\ and\ \citenamefont {Ollitrault}}]{Gardim12a}%
  \BibitemOpen
  \bibfield  {author} {\bibinfo {author} {\bibfnamefont {F.~G.}\ \bibnamefont
  {Gardim}}, \bibinfo {author} {\bibfnamefont {F.}~\bibnamefont {Grassi}},
  \bibinfo {author} {\bibfnamefont {M.}~\bibnamefont {Luzum}}, \ and\ \bibinfo
  {author} {\bibfnamefont {J.-Y.}\ \bibnamefont {Ollitrault}},\ }\href@noop {}
  {\bibfield  {journal} {\bibinfo  {journal} {Phys. Rev. Lett.}\ }\textbf
  {\bibinfo {volume} {109}},\ \bibinfo {pages} {202302} (\bibinfo {year}
  {2012}{\natexlab{b}})},\ \Eprint {http://arxiv.org/abs/arXiv:1203.2882}
  {arXiv:1203.2882} \BibitemShut {NoStop}%
\bibitem [{\citenamefont {Takahashi}\ \emph {et~al.}(2009)\citenamefont
  {Takahashi} \emph {et~al.}}]{Takahashi09}%
  \BibitemOpen
  \bibfield  {author} {\bibinfo {author} {\bibfnamefont {J.}~\bibnamefont
  {Takahashi}} \emph {et~al.},\ }\href@noop {} {\bibfield  {journal} {\bibinfo
  {journal} {Phys. Rev. Lett.}\ }\textbf {\bibinfo {volume} {103}},\ \bibinfo
  {pages} {242301} (\bibinfo {year} {2009})},\ \Eprint
  {http://arxiv.org/abs/arXiv:0902.4870} {arXiv:0902.4870} \BibitemShut
  {NoStop}%
\bibitem [{\citenamefont {W.L.Qian}\ \emph {et~al.}(2013)\citenamefont
  {W.L.Qian}, \citenamefont {Andrade}, \citenamefont {Gardim}, \citenamefont
  {Grassi},\ and\ \citenamefont {Hama}}]{Qian12}%
  \BibitemOpen
  \bibfield  {author} {\bibinfo {author} {\bibnamefont {W.L.Qian}}, \bibinfo
  {author} {\bibfnamefont {R.~P.~G.}\ \bibnamefont {Andrade}}, \bibinfo
  {author} {\bibfnamefont {F.}~\bibnamefont {Gardim}}, \bibinfo {author}
  {\bibfnamefont {F.}~\bibnamefont {Grassi}}, \ and\ \bibinfo {author}
  {\bibfnamefont {Y.}~\bibnamefont {Hama}},\ }\href@noop {} {\bibfield
  {journal} {\bibinfo  {journal} {Phys. Rev. C}\ }\textbf {\bibinfo {volume}
  {87}},\ \bibinfo {pages} {014904} (\bibinfo {year} {2013})},\ \Eprint
  {http://arxiv.org/abs/arXiv:1207.6415} {arXiv:1207.6415} \BibitemShut
  {NoStop}%
\bibitem [{\citenamefont {Machado}(2015)}]{diss_meera}%
  \BibitemOpen
  \bibfield  {author} {\bibinfo {author} {\bibfnamefont {M.~V.}\ \bibnamefont
  {Machado}},\ }\emph {\bibinfo {title} {Event-by-event hydrodynamics for the
  LHC}},\ \href@noop {} {Master's thesis},\ \bibinfo  {school} {Universidade de
  S\~ao Paulo}, \bibinfo {address} {Brazil} (\bibinfo {year}
  {2015})\BibitemShut {NoStop}%
\bibitem [{\citenamefont {Gardim}\ \emph {et~al.}(2020)\citenamefont {Gardim},
  \citenamefont {Grassi}, \citenamefont {Ishida}, \citenamefont {Luzum},\ and\
  \citenamefont {Ollitrault}}]{Gardim:2020fxx}%
  \BibitemOpen
  \bibfield  {author} {\bibinfo {author} {\bibfnamefont {F.~G.}\ \bibnamefont
  {Gardim}}, \bibinfo {author} {\bibfnamefont {F.}~\bibnamefont {Grassi}},
  \bibinfo {author} {\bibfnamefont {P.}~\bibnamefont {Ishida}}, \bibinfo
  {author} {\bibfnamefont {M.}~\bibnamefont {Luzum}}, \ and\ \bibinfo {author}
  {\bibfnamefont {J.-Y.}\ \bibnamefont {Ollitrault}},\ }in\ \href@noop {}
  {\emph {\bibinfo {booktitle} {{28th International Conference on
  Ultrarelativistic Nucleus-Nucleus Collisions}}}}\ (\bibinfo {year} {2020})\
  \Eprint {http://arxiv.org/abs/2002.01747} {arXiv:2002.01747 [nucl-th]}
  \BibitemShut {NoStop}%
\bibitem [{\citenamefont {Bernhard}\ \emph {et~al.}(2016)\citenamefont
  {Bernhard}, \citenamefont {Moreland}, \citenamefont {Bass}, \citenamefont
  {Liu},\ and\ \citenamefont {Heinz}}]{Bernhard:2016tnd}%
  \BibitemOpen
  \bibfield  {author} {\bibinfo {author} {\bibfnamefont {J.~E.}\ \bibnamefont
  {Bernhard}}, \bibinfo {author} {\bibfnamefont {J.~S.}\ \bibnamefont
  {Moreland}}, \bibinfo {author} {\bibfnamefont {S.~A.}\ \bibnamefont {Bass}},
  \bibinfo {author} {\bibfnamefont {J.}~\bibnamefont {Liu}}, \ and\ \bibinfo
  {author} {\bibfnamefont {U.}~\bibnamefont {Heinz}},\ }\href {\doibase
  10.1103/PhysRevC.94.024907} {\bibfield  {journal} {\bibinfo  {journal} {Phys.
  Rev.}\ }\textbf {\bibinfo {volume} {C94}},\ \bibinfo {pages} {024907}
  (\bibinfo {year} {2016})},\ \Eprint {http://arxiv.org/abs/1605.03954}
  {arXiv:1605.03954 [nucl-th]} \BibitemShut {NoStop}%
%%CITATION = ARXIV:1605.03954;%%
\bibitem [{\citenamefont {Alba}\ \emph
  {et~al.}(2017{\natexlab{a}})\citenamefont {Alba}, \citenamefont {Sarti},
  \citenamefont {Noronha}, \citenamefont {Noronha-Hostler}, \citenamefont
  {Parotto}, \citenamefont {Vazquez},\ and\ \citenamefont
  {Ratti}}]{Alba:2017hhe}%
  \BibitemOpen
  \bibfield  {author} {\bibinfo {author} {\bibfnamefont {P.}~\bibnamefont
  {Alba}}, \bibinfo {author} {\bibfnamefont {V.~M.}\ \bibnamefont {Sarti}},
  \bibinfo {author} {\bibfnamefont {J.}~\bibnamefont {Noronha}}, \bibinfo
  {author} {\bibfnamefont {J.}~\bibnamefont {Noronha-Hostler}}, \bibinfo
  {author} {\bibfnamefont {P.}~\bibnamefont {Parotto}}, \bibinfo {author}
  {\bibfnamefont {I.~P.}\ \bibnamefont {Vazquez}}, \ and\ \bibinfo {author}
  {\bibfnamefont {C.}~\bibnamefont {Ratti}},\ }\href@noop {} {\  (\bibinfo
  {year} {2017}{\natexlab{a}})},\ \Eprint {http://arxiv.org/abs/1711.05207}
  {arXiv:1711.05207 [nucl-th]} \BibitemShut {NoStop}%
%%CITATION = ARXIV:1711.05207;%%
\bibitem [{\citenamefont {Nijs}\ \emph {et~al.}(2020)\citenamefont {Nijs},
  \citenamefont {Van Der~Schee}, \citenamefont {G\"ursoy},\ and\ \citenamefont
  {Snellings}}]{Nijs:2020roc}%
  \BibitemOpen
  \bibfield  {author} {\bibinfo {author} {\bibfnamefont {G.}~\bibnamefont
  {Nijs}}, \bibinfo {author} {\bibfnamefont {W.}~\bibnamefont {Van Der~Schee}},
  \bibinfo {author} {\bibfnamefont {U.}~\bibnamefont {G\"ursoy}}, \ and\
  \bibinfo {author} {\bibfnamefont {R.}~\bibnamefont {Snellings}},\ }\href@noop
  {} {\  (\bibinfo {year} {2020})},\ \Eprint {http://arxiv.org/abs/2010.15134}
  {arXiv:2010.15134 [nucl-th]} \BibitemShut {NoStop}%
\bibitem [{\citenamefont {Summerfield}\ \emph {et~al.}(2021)\citenamefont
  {Summerfield}, \citenamefont {Lu}, \citenamefont {Plumberg}, \citenamefont
  {Lee}, \citenamefont {Noronha-Hostler},\ and\ \citenamefont
  {Timmins}}]{Summerfield:2021oex}%
  \BibitemOpen
  \bibfield  {author} {\bibinfo {author} {\bibfnamefont {N.}~\bibnamefont
  {Summerfield}}, \bibinfo {author} {\bibfnamefont {B.-N.}\ \bibnamefont {Lu}},
  \bibinfo {author} {\bibfnamefont {C.}~\bibnamefont {Plumberg}}, \bibinfo
  {author} {\bibfnamefont {D.}~\bibnamefont {Lee}}, \bibinfo {author}
  {\bibfnamefont {J.}~\bibnamefont {Noronha-Hostler}}, \ and\ \bibinfo {author}
  {\bibfnamefont {A.}~\bibnamefont {Timmins}},\ }\href@noop {} {\  (\bibinfo
  {year} {2021})},\ \Eprint {http://arxiv.org/abs/2103.03345} {arXiv:2103.03345
  [nucl-th]} \BibitemShut {NoStop}%
\bibitem [{\citenamefont {Ke}\ \emph {et~al.}(2017)\citenamefont {Ke},
  \citenamefont {Moreland}, \citenamefont {Bernhard},\ and\ \citenamefont
  {Bass}}]{Ke:2016jrd}%
  \BibitemOpen
  \bibfield  {author} {\bibinfo {author} {\bibfnamefont {W.}~\bibnamefont
  {Ke}}, \bibinfo {author} {\bibfnamefont {J.~S.}\ \bibnamefont {Moreland}},
  \bibinfo {author} {\bibfnamefont {J.~E.}\ \bibnamefont {Bernhard}}, \ and\
  \bibinfo {author} {\bibfnamefont {S.~A.}\ \bibnamefont {Bass}},\ }\href
  {\doibase 10.1103/PhysRevC.96.044912} {\bibfield  {journal} {\bibinfo
  {journal} {Phys. Rev. C}\ }\textbf {\bibinfo {volume} {96}},\ \bibinfo
  {pages} {044912} (\bibinfo {year} {2017})},\ \Eprint
  {http://arxiv.org/abs/1610.08490} {arXiv:1610.08490 [nucl-th]} \BibitemShut
  {NoStop}%
\bibitem [{\citenamefont {Mordasini}\ \emph {et~al.}(2020)\citenamefont
  {Mordasini}, \citenamefont {Bilandzic}, \citenamefont {Karako\c{c}},\ and\
  \citenamefont {Taghavi}}]{Mordasini:2019hut}%
  \BibitemOpen
  \bibfield  {author} {\bibinfo {author} {\bibfnamefont {C.}~\bibnamefont
  {Mordasini}}, \bibinfo {author} {\bibfnamefont {A.}~\bibnamefont
  {Bilandzic}}, \bibinfo {author} {\bibfnamefont {D.}~\bibnamefont
  {Karako\c{c}}}, \ and\ \bibinfo {author} {\bibfnamefont {S.~F.}\ \bibnamefont
  {Taghavi}},\ }\href {\doibase 10.1103/PhysRevC.102.024907} {\bibfield
  {journal} {\bibinfo  {journal} {Phys. Rev. C}\ }\textbf {\bibinfo {volume}
  {102}},\ \bibinfo {pages} {024907} (\bibinfo {year} {2020})},\ \Eprint
  {http://arxiv.org/abs/1901.06968} {arXiv:1901.06968 [nucl-ex]} \BibitemShut
  {NoStop}%
\bibitem [{\citenamefont {Giacalone}\ \emph
  {et~al.}(2021{\natexlab{a}})\citenamefont {Giacalone}, \citenamefont
  {Gardim}, \citenamefont {Noronha-Hostler},\ and\ \citenamefont
  {Ollitrault}}]{Giacalone:2020lbm}%
  \BibitemOpen
  \bibfield  {author} {\bibinfo {author} {\bibfnamefont {G.}~\bibnamefont
  {Giacalone}}, \bibinfo {author} {\bibfnamefont {F.~G.}\ \bibnamefont
  {Gardim}}, \bibinfo {author} {\bibfnamefont {J.}~\bibnamefont
  {Noronha-Hostler}}, \ and\ \bibinfo {author} {\bibfnamefont {J.-Y.}\
  \bibnamefont {Ollitrault}},\ }\href {\doibase 10.1103/PhysRevC.103.024910}
  {\bibfield  {journal} {\bibinfo  {journal} {Phys. Rev. C}\ }\textbf {\bibinfo
  {volume} {103}},\ \bibinfo {pages} {024910} (\bibinfo {year}
  {2021}{\natexlab{a}})},\ \Eprint {http://arxiv.org/abs/2004.09799}
  {arXiv:2004.09799 [nucl-th]} \BibitemShut {NoStop}%
\bibitem [{\citenamefont {Giacalone}\ \emph
  {et~al.}(2021{\natexlab{b}})\citenamefont {Giacalone}, \citenamefont
  {Gardim}, \citenamefont {Noronha-Hostler},\ and\ \citenamefont
  {Ollitrault}}]{Giacalone:2020dln}%
  \BibitemOpen
  \bibfield  {author} {\bibinfo {author} {\bibfnamefont {G.}~\bibnamefont
  {Giacalone}}, \bibinfo {author} {\bibfnamefont {F.~G.}\ \bibnamefont
  {Gardim}}, \bibinfo {author} {\bibfnamefont {J.}~\bibnamefont
  {Noronha-Hostler}}, \ and\ \bibinfo {author} {\bibfnamefont {J.-Y.}\
  \bibnamefont {Ollitrault}},\ }\href {\doibase 10.1103/PhysRevC.103.024909}
  {\bibfield  {journal} {\bibinfo  {journal} {Phys. Rev. C}\ }\textbf {\bibinfo
  {volume} {103}},\ \bibinfo {pages} {024909} (\bibinfo {year}
  {2021}{\natexlab{b}})},\ \Eprint {http://arxiv.org/abs/2004.01765}
  {arXiv:2004.01765 [nucl-th]} \BibitemShut {NoStop}%
\bibitem [{\citenamefont {ATLAS}(2021)}]{ATLAS:2021kty}%
  \BibitemOpen
  \bibfield  {author} {\bibinfo {author} {\bibnamefont {ATLAS}},\ }\href@noop
  {} {\enquote {\bibinfo {title} {{Measurement of flow and transverse momentum
  correlations in Pb+Pb collisions at $\sqrt{s_{\mathrm{NN}}}=5.02$ TeV and
  Xe+Xe collisions at $\sqrt{s_{\mathrm{NN}}}=5.44$ TeV with the ATLAS
  detector}},}\ } (\bibinfo {year} {2021})\BibitemShut {NoStop}%
\bibitem [{\citenamefont {Bernhard}\ \emph {et~al.}(2019)\citenamefont
  {Bernhard}, \citenamefont {Moreland},\ and\ \citenamefont
  {Bass}}]{Bernhard:2019bmu}%
  \BibitemOpen
  \bibfield  {author} {\bibinfo {author} {\bibfnamefont {J.~E.}\ \bibnamefont
  {Bernhard}}, \bibinfo {author} {\bibfnamefont {J.~S.}\ \bibnamefont
  {Moreland}}, \ and\ \bibinfo {author} {\bibfnamefont {S.~A.}\ \bibnamefont
  {Bass}},\ }\href {\doibase 10.1038/s41567-019-0611-8} {\bibfield  {journal}
  {\bibinfo  {journal} {Nature Phys.}\ }\textbf {\bibinfo {volume} {15}},\
  \bibinfo {pages} {1113} (\bibinfo {year} {2019})}\BibitemShut {NoStop}%
\bibitem [{\citenamefont {Aguiar}\ \emph {et~al.}(2001)\citenamefont {Aguiar},
  \citenamefont {Kodama}, \citenamefont {Osada},\ and\ \citenamefont
  {Hama}}]{testNeX}%
  \BibitemOpen
  \bibfield  {author} {\bibinfo {author} {\bibfnamefont {C.}~\bibnamefont
  {Aguiar}}, \bibinfo {author} {\bibfnamefont {T.}~\bibnamefont {Kodama}},
  \bibinfo {author} {\bibfnamefont {T.}~\bibnamefont {Osada}}, \ and\ \bibinfo
  {author} {\bibfnamefont {Y.}~\bibnamefont {Hama}},\ }\href@noop {} {\bibfield
   {journal} {\bibinfo  {journal} {J.Phys.G}\ }\textbf {\bibinfo {volume}
  {27}},\ \bibinfo {pages} {75} (\bibinfo {year} {2001})},\ \Eprint
  {http://arxiv.org/abs/arXiv:hep-ph/0006239} {arXiv:hep-ph/0006239}
  \BibitemShut {NoStop}%
\bibitem [{\citenamefont {Hama}\ \emph {et~al.}(2006)\citenamefont {Hama},
  \citenamefont {Andrade}, \citenamefont {Grassi}, \citenamefont {Jr},
  \citenamefont {Kodama}, \citenamefont {Tavares},\ and\ \citenamefont
  {Padula}}]{eos}%
  \BibitemOpen
  \bibfield  {author} {\bibinfo {author} {\bibfnamefont {Y.}~\bibnamefont
  {Hama}}, \bibinfo {author} {\bibfnamefont {R.~P.}\ \bibnamefont {Andrade}},
  \bibinfo {author} {\bibfnamefont {F.}~\bibnamefont {Grassi}}, \bibinfo
  {author} {\bibfnamefont {O.~S.}\ \bibnamefont {Jr}}, \bibinfo {author}
  {\bibfnamefont {T.}~\bibnamefont {Kodama}}, \bibinfo {author} {\bibfnamefont
  {B.}~\bibnamefont {Tavares}}, \ and\ \bibinfo {author} {\bibfnamefont
  {S.~S.}\ \bibnamefont {Padula}},\ }\href@noop {} {\bibfield  {journal}
  {\bibinfo  {journal} {Nucl.Phys.A}\ }\textbf {\bibinfo {volume} {169}},\
  \bibinfo {pages} {774} (\bibinfo {year} {2006})},\ \Eprint
  {http://arxiv.org/abs/arXiv:hep-ph/0510096} {arXiv:hep-ph/0510096}
  \BibitemShut {NoStop}%
\bibitem [{\citenamefont {Noronha-Hostler}\ \emph {et~al.}(2013)\citenamefont
  {Noronha-Hostler}, \citenamefont {Denicol}, \citenamefont {Noronha},
  \citenamefont {Andrade},\ and\ \citenamefont {Grassi}}]{vuspb}%
  \BibitemOpen
  \bibfield  {author} {\bibinfo {author} {\bibfnamefont {J.}~\bibnamefont
  {Noronha-Hostler}}, \bibinfo {author} {\bibfnamefont {G.~S.}\ \bibnamefont
  {Denicol}}, \bibinfo {author} {\bibfnamefont {J.}~\bibnamefont {Noronha}},
  \bibinfo {author} {\bibfnamefont {R.~P.~G.}\ \bibnamefont {Andrade}}, \ and\
  \bibinfo {author} {\bibfnamefont {F.}~\bibnamefont {Grassi}},\ }\href@noop {}
  {\bibfield  {journal} {\bibinfo  {journal} {Phys. Rev. C}\ }\textbf {\bibinfo
  {volume} {88}},\ \bibinfo {pages} {044916} (\bibinfo {year} {2013})},\
  \Eprint {http://arxiv.org/abs/arXiv:1305.1981} {arXiv:1305.1981} \BibitemShut
  {NoStop}%
\bibitem [{\citenamefont {Noronha-Hostler}\ \emph {et~al.}(2014)\citenamefont
  {Noronha-Hostler}, \citenamefont {Noronha},\ and\ \citenamefont
  {Grassi}}]{vuspbs}%
  \BibitemOpen
  \bibfield  {author} {\bibinfo {author} {\bibfnamefont {J.}~\bibnamefont
  {Noronha-Hostler}}, \bibinfo {author} {\bibfnamefont {J.}~\bibnamefont
  {Noronha}}, \ and\ \bibinfo {author} {\bibfnamefont {F.}~\bibnamefont
  {Grassi}},\ }\href@noop {} {\bibfield  {journal} {\bibinfo  {journal} {Phys.
  Rev. C}\ }\textbf {\bibinfo {volume} {90}},\ \bibinfo {pages} {034907}
  (\bibinfo {year} {2014})},\ \Eprint {http://arxiv.org/abs/arXiv:1406.3333}
  {arXiv:1406.3333} \BibitemShut {NoStop}%
\bibitem [{\citenamefont {Alba}\ \emph
  {et~al.}(2017{\natexlab{b}})\citenamefont {Alba} \emph
  {et~al.}}]{Alba:2017mqu}%
  \BibitemOpen
  \bibfield  {author} {\bibinfo {author} {\bibfnamefont {P.}~\bibnamefont
  {Alba}} \emph {et~al.},\ }\href {\doibase 10.1103/PhysRevD.96.034517}
  {\bibfield  {journal} {\bibinfo  {journal} {Phys. Rev. D}\ }\textbf {\bibinfo
  {volume} {96}},\ \bibinfo {pages} {034517} (\bibinfo {year}
  {2017}{\natexlab{b}})},\ \Eprint {http://arxiv.org/abs/1702.01113}
  {arXiv:1702.01113 [hep-lat]} \BibitemShut {NoStop}%
\bibitem [{\citenamefont {Song}\ and\ \citenamefont {Heinz}(2008)}]{Song_2008}%
  \BibitemOpen
  \bibfield  {author} {\bibinfo {author} {\bibfnamefont {H.}~\bibnamefont
  {Song}}\ and\ \bibinfo {author} {\bibfnamefont {U.}~\bibnamefont {Heinz}},\
  }\href {\doibase 10.1103/physrevc.77.064901} {\bibfield  {journal} {\bibinfo
  {journal} {Physical Review C}\ }\textbf {\bibinfo {volume} {77}} (\bibinfo
  {year} {2008}),\ 10.1103/physrevc.77.064901}\BibitemShut {NoStop}%
\bibitem [{\citenamefont {Song}\ \emph {et~al.}(2011)\citenamefont {Song},
  \citenamefont {Bass},\ and\ \citenamefont {Heinz}}]{Song_2011}%
  \BibitemOpen
  \bibfield  {author} {\bibinfo {author} {\bibfnamefont {H.}~\bibnamefont
  {Song}}, \bibinfo {author} {\bibfnamefont {S.~A.}\ \bibnamefont {Bass}}, \
  and\ \bibinfo {author} {\bibfnamefont {U.}~\bibnamefont {Heinz}},\ }\href
  {\doibase 10.1103/physrevc.83.024912} {\bibfield  {journal} {\bibinfo
  {journal} {Physical Review C}\ }\textbf {\bibinfo {volume} {83}} (\bibinfo
  {year} {2011}),\ 10.1103/physrevc.83.024912}\BibitemShut {NoStop}%
\bibitem [{\citenamefont {Giacalone}\ \emph
  {et~al.}(2017{\natexlab{b}})\citenamefont {Giacalone}, \citenamefont {Yan},
  \citenamefont {Noronha-Hostler},\ and\ \citenamefont
  {Ollitrault}}]{Giacalone:2016mdr}%
  \BibitemOpen
  \bibfield  {author} {\bibinfo {author} {\bibfnamefont {G.}~\bibnamefont
  {Giacalone}}, \bibinfo {author} {\bibfnamefont {L.}~\bibnamefont {Yan}},
  \bibinfo {author} {\bibfnamefont {J.}~\bibnamefont {Noronha-Hostler}}, \ and\
  \bibinfo {author} {\bibfnamefont {J.-Y.}\ \bibnamefont {Ollitrault}},\ }\href
  {\doibase 10.1088/1742-6596/779/1/012064} {\bibfield  {journal} {\bibinfo
  {journal} {J. Phys. Conf. Ser.}\ }\textbf {\bibinfo {volume} {779}},\
  \bibinfo {pages} {012064} (\bibinfo {year} {2017}{\natexlab{b}})},\ \Eprint
  {http://arxiv.org/abs/1608.06022} {arXiv:1608.06022 [nucl-th]} \BibitemShut
  {NoStop}%
\end{thebibliography}%

\end{document}